\RequirePackage{silence}
\documentclass[journal=tches,floatrow,preprint]{iacrtrans}

\usepackage{lipsum}
\usepackage{subfiles}
\usepackage[ruled,noline]{algorithm2e}
\SetProcNameSty{textsc}

\usepackage[mathscr]{euscript}
\usepackage{multirow}
\usepackage{makecell}
\usepackage{subcaption}
\usepackage{mdframed}
\usepackage{multicol}
\usepackage{array}
\usepackage{tabularx}
\usepackage{comment}
\usepackage{listings}
\usepackage{wrapfig}
\usepackage{anyfontsize}
\usepackage{lstlangarm} 
\usepackage{pgfplots}
\usepgfplotslibrary{fillbetween,groupplots}
\pgfplotsset{compat=1.18}

\usepackage{longtable}
\usepackage{pifont} 
\newcommand{\cmark}{\ding{51}} 
\newcommand{\xmark}{\ding{55}} 

\usepackage{tikz}
\usepackage{todonotes}

\usepackage{arydshln} 
\usepackage{amsthm} 
\usepackage[inline,shortlabels]{enumitem}
\usepackage[htt]{hyphenat}
\usepackage{esvect}

\usepackage{circuitikz}
\ctikzset{logic ports=ieee, logic ports/scale=0.75}

\usetikzlibrary{positioning, intersections, shapes, arrows.meta, math, bending, calc, backgrounds, fit, decorations.pathmorphing}

\newcommand{\framework}{aLEAKator}

\newcommand{\tool}[1]{\texttt{#1}}
\newcommand{\vsec}{\texttt{sec-v}}
\newcommand{\vsecrr}{\texttt{sec-rr}}
\newcommand{\xsec}{\texttt{nsec}}
\newcommand{\lmd}[1]{\texttt{#1}}
\newcommand{\token}[1]{\ensuremath{\mathtt{#1}}}
\newcommand{\coco}{\tool{Coco}}
\newcommand{\prolead}{\tool{Prolead}}
\newcommand{\prover}{\tool{Prover}}

\newcommand{\cmthree}{Cortex-M3}
\newcommand{\cmfour}{Cortex-M4}
\newcommand{\ibex}{Ibex}
\newcommand{\cocoibex}{Coco-Ibex}
\newcommand{\cv}{CV32E40P}

\colorlet{clset}{blue}
\colorlet{cstab}{violet}
\definecolor{cconc}{RGB}{69, 18, 0}
\colorlet{csymb}{teal}

\colorlet{cleak}{red}
\definecolor{cnoleak}{RGB}{0, 128, 0}

\newcommand{\dconc}{\ensuremath{\mathbb{B}}}
\newcommand{\dsymb}{\ensuremath{\mathbb{E}}}
\newcommand{\dlset}{\ensuremath{\mathbb{L}}}
\newcommand{\dstab}{\ensuremath{\mathbb{S}}}

\newcommand{\fconc}[1]{\ensuremath{\textcolor{cconc}{{}^{c}#1}}}
\newcommand{\fsymb}[1]{\ensuremath{\textcolor{csymb}{{}^{e}#1}}}
\newcommand{\flset}[1]{\ensuremath{\textcolor{clset}{{}^{l}#1}}}
\newcommand{\fstab}[1]{\ensuremath{\textcolor{cstab}{{}^{s}#1}}}

\newcommand{\vconc}[2][]{\ensuremath{\textcolor{cconc}{{}^{c}\alpha^{#1}_{#2}}}}
\newcommand{\vsymb}[2][]{\ensuremath{\textcolor{csymb}{{}^{e}\alpha^{#1}_{#2}}}}
\newcommand{\vlset}[2][]{\ensuremath{\textcolor{clset}{{}^{l}\alpha^{#1}_{#2}}}}
\newcommand{\vstab}[2][]{\ensuremath{\textcolor{cstab}{{}^{s}\alpha^{#1}_{#2}}}}

\newcommand{\ls}{\textit{LeakSet}}
\newcommand{\lss}{\textit{LeakSets}}

\newcommand*\circled[1]{\tikz[baseline=(char.base)] \node[draw, circle, font=\small, inner sep=0.25mm] (char) {#1};}

\definecolor{Paired-2}{RGB}{166,206,227}
\definecolor{Paired-1}{RGB}{31,120,180}
\definecolor{Paired-4}{RGB}{178,223,138}
\definecolor{Paired-3}{RGB}{51,160,44}
\definecolor{Paired-6}{RGB}{251,154,153}
\definecolor{Paired-5}{RGB}{227,26,28}
\definecolor{Paired-8}{RGB}{253,191,111}
\definecolor{Paired-7}{RGB}{255,127,0}
\definecolor{Paired-10}{RGB}{202,178,214}
\definecolor{Paired-9}{RGB}{106,61,154}
\definecolor{Paired-12}{RGB}{255,255,153}
\definecolor{Paired-11}{RGB}{177,89,40}
\definecolor{Accent-1}{RGB}{127,201,127}
\definecolor{Accent-2}{RGB}{190,174,212}
\definecolor{Accent-3}{RGB}{253,192,134}
\definecolor{Accent-4}{RGB}{255,255,153}
\definecolor{Accent-5}{RGB}{56,108,176}
\definecolor{Accent-6}{RGB}{240,2,127}
\definecolor{Accent-7}{RGB}{191,91,23}
\definecolor{Accent-8}{RGB}{102,102,102}

\colorlet{cmethod}{Paired-2}
\colorlet{ccpuspec}{Accent-5}

\author{Noé Amiot\texorpdfstring{\inst{1}}{} \and Quentin Meunier\texorpdfstring{\inst{1}}{} \and Karine Heydemann\texorpdfstring{\inst{1,2}}{} \and Emmanuelle Encrenaz\texorpdfstring{\inst{1}}{}}
\institute{
  Sorbonne Université, LIP6, CNRS, 4 Place Jussieu 75005, Paris, France, \email{first.last@lip6.fr}
  \and
  Thales, Meyreuil, France, \email{first.last@thalesgroup.com}
}

\title[\framework{}: Mixed-Domain Simulation for Masked Implementations Verification]{\framework{}: HDL Mixed-Domain Simulation for Masked Hardware\hspace{0.15em}\&\hspace{0.15em}Software Formal Verification}

\begin{document}
\maketitle


\begin{abstract}
Verifying the security of masked hardware and software implementations, under advanced leakage models, remains a significant challenge, especially when accounting for glitches, transitions and CPU micro-architectural specifics. Existing verification approaches are either restricted to small hardware gadgets, small programs on CPUs such as Sboxes, limited leakage models, or require hardware-specific prior knowledge.

In this work, we present \framework{}, an open-source framework for the automated formal verification of masked cryptographic accelerators and software running on CPUs from their HDL descriptions. Our method introduces mixed-domain simulation, enabling precise modeling and verification under various (including robust and relaxed) 1-probing leakage models, and supports variable signal granularity without being restricted to 1-bit wires. \framework{} also supports verification in the presence of lookup tables, and does not require prior knowledge of the target CPU architecture. Our approach is validated against existing tools and real-world measurements while providing innovative results such as the verification of a full, first-order masked AES on various CPUs.
\end{abstract}

\section{Introduction}
\label{sc:introduction}

\subsection{Context}

Side-channel attacks, discovered in the late 90's~\cite{C:Kocher96,C:KocJafJun99}, are still a major threat for hardware and software implementations of cryptographic algorithms. These attacks exploit physical quantities such as power consumption, electromagnetic emissions, or timing information to retrieve secret data manipulated by a cryptographic system.

Introduced in 1999 \cite{CHES:GouPat99,C:CJRR99}, the \emph{masking} countermeasure has been formalised by Ishai \textit{et al.} \cite{C:IshSahWag03}. 
At order $d$, it consists in splitting a secret variable $s$ into $d+1$ parts $s_0 \dots s_d$, called \emph{shares}, using $d$ uniformly distributed random values and such that $s = s_0 \star \dots \star s_{d}$ where $\star$ is the operator used for sharing. For example, Boolean masking uses the $\oplus$ operator, whereas Arithmetic masking uses the arithmetic addition $+$. With such a sharing, any combination of $d$ or less shares is statistically independent of the secret variable $s$. Since the seminal Boolean masking scheme~\cite{C:IshSahWag03}, other schemes such as Threshold Implementation (TI)~\cite{ICICS:NikRecRij06} and Domain-Oriented Masking (DOM)~\cite{grossDomainOrientedMaskingCompact2016} have been proposed to either eliminate the randomness needs, resist glitches or even ease the secure composability of masked circuits.

A masked implementation can be proven secure according to a \emph{leakage model} which defines the information an attacker can retrieve while observing the circuit or the running program. In the \lmd{d-probing model}~\cite{C:IshSahWag03}, an attacker is able to observe any \lmd{d}-uplet of intermediate values computed by the implementation. An implementation is said \lmd{d-probing secure} if any \lmd{d}-uplet is statistically independent of secret data. Yet, this rather simple and widely used model does not always match real leakages, as other physical effects contribute to the dissipated power consumption or EM emanations. Due to the propagation delay of signals through gates and wires, the input wires of a gate may stabilise at different instants or change several times before stabilising, hence inducing a potential transient toggling of the gate output, named \emph{glitch}, which possibly reveals information on the inputs of the gate. When using the CMOS technology, significant instantaneous power consumption occurs when a wire changes from a logic value 0 to a logic value 1 or conversely~\cite{Mangard2007PowerAA}. As a consequence, an attacker is able to observe the \emph{transitions} between consecutive values of a wire with a single probe. Extended leakage models have been introduced to cover these hardware effects. The \lmd{robust d-probing model}~\cite{TCHES:FGMPS18} extends the \lmd{d-probing model} with either transitions, glitches, or both, and is expressed as \lmd{(g,t) d-probing model}\footnote{We omit the parameter \lmd{c} for the coupling effects, not accounted for in the following.}, in which \lmd{g} and \lmd{t} are set to 0 or 1 accordingly.

In order to reduce the set of considered glitches only to those that may occur in practice, the notion of signal \emph{stability} was recently introduced when glitches cannot occur~\cite{CocoAlma,proleadv3}. Specifically, the stability of a wire indicates whether its value has changed during the current cycle. For a register, its output is stable if and only if its value is the same as in the previous cycle. For a combinatorial gate, its output is stable when its whole input combinatorial dependency tree does not change value in the current cycle. As stable wires cannot induce glitches, a probe on a stable wire only leaks the wire value for the current cycle. The \lmd{Robust but Relaxed d-probing model}~\cite{proleadv3} or \lmd{RR d-probing model} is defined as the \lmd{(1,1) d-probing model} while considering stability of the signals.

\subsection{Open Research Questions}

Various formal approaches have been proposed to prove whether a masked hardware or
software implementation leaks within a specified leakage model. Verifying the absence of secret leakage in a software masked implementation requires to consider the target device micro-architecture that dictates the potential source of leakages hence reducing the security of the masked software~\cite{TCHES:MarPagWeb22, ProvableMaskingInRealWorld,AC:GigPriMan21}. A potential source of leakage causes a leak depending on the code being executed, advocating for a co-verification of the hardware and the software. Moreover, if the composition of formally verified masked gadgets is a possible approach for producing a fully masked implementation, it requires formally proving the security of gadget composition. For hardware implementations, this is already covered by several works proposing security properties covering secure composition such as Strong Non-Interference (SNI)~\cite{CCS:BBDFGS16} and Probe Isolating Non-Interference (PINI)~\cite{cassiers2020trivially}, as well as methods for formally proving them~\cite{ESORICS:BBCFGS19,AC:KniSasMor20,prover}. Regarding software, secure composition of gadgets is still an open question, in addition to the fact that masking schemes dedicated to a specific algorithm are more likely to be deployed, as they most often simultaneously reduce the required amount of randomness, the execution time, and the code size w.r.t. a gadget-based masked implementation.
Today, state-of-the-art masked software verification solutions are however limited to 1) either small programs, typically software gadgets, with a limited number of executed cycles while possibly considering the most powerful leakage model \cite{USENIX:GHPMB21,AC:GigPriMan21}; 2) fully masked programs while only considering \lmd{(0,1) 1-probing leakage model} and a manually extracted model from the CPU design~\cite{armistice}, potentially leading to missed leakage sources; 3) full program verification using a 2-step approach based on contracts \cite{closingthegap}. Contracts must be verified once against the hardware target, then the verification of the software is performed considering the verified contracts. Such power contracts are manually defined, necessitating advanced prior knowledge of the target device.

As some optimised masked implementations make use of lookup tables to retrieve precomputed results, e.g.~\cite{ACNS:HerOswMan06}, verification techniques should be able to support such masking optimisations. Verifying a masked software running on a CPU model resulting from the synthesis step --- i.e. once the mapping with the target technology has been carried out --- allows to only detect leakages visible after this mapping but it has two drawbacks. First, the verification output is technology-dependent; second, the verification is necessarily carried out at the bit level~\cite{USENIX:GHPMB21}. At this level, it is difficult to identify all the bits composing a wire after the synthesis.
It is therefore not possible to correctly verify masked implementations deployed using $n$-bit shares because it becomes difficult to recombine $n$-bit wide intermediate values, when $n$ is greater that one.


As a result, overcoming the limitations identified in current verification approaches remains an open challenge. A key objective is to enable the automatic verification of larger programs running on a wide range of CPUs, under strong leakage models. In particular, achieving verification without requiring prior knowledge of the target CPU architecture, and without restricting the analysis to 1-bit wires, is a crucial property of future verification techniques.

\subsection{Contributions}

To address the aforementioned challenges, we propose a novel methodology, implemented as a dedicated tool, that enables the verification of the absence of secret leakage in both cryptographic hardware accelerators and masked software executing on CPUs.

Specifically, we introduce and formally define a new simulation technique called \emph{mixed-domain} simulation. This approach allows the comprehensive collection of all necessary information to perform verification within a range of leakage models, including the \lmd{RR 1-probing model}. Furthermore, we systematically identify and specify the set of wires that require verification for each supported leakage model. This targeted selection significantly reduces the number of verifications required to assess a property, without compromising security: any potential leakage from a non-verified signal necessarily propagates to a verified signal.

We present \framework{}, implementing the mixed-domain simulation technique and verification requests optimisation. \framework{} is a comprehensive framework for the automated formal verification of cryptographic accelerators and programs running on CPUs: It processes the HDL code of a masked hardware or of a system-on-chip embedding a CPU to extract a model representative of post-synthesis designs, preserving original signal widths and that can be simulated. This enables for the verification at varying levels of granularity, whether at the bit or signal level. The framework supports the specification of tailored verification rules for masked programs involving lookup table accesses. Notably, \framework{} is open source\footnote{Available at \url{https://github.com/noeamiot/aLEAKator}}, with all necessary tooling for model extraction and verification.

To demonstrate the versatility, efficiency, and accuracy of \framework{}, we conduct extensive experiments. First, we validate our method by successfully reproducing eight state-of-the-art formal verifications of cryptographic hardware, achieving up to a 50-fold speed-up over previous approaches. Next, we assess the security of both the original and a hardened version of a masked gadget, as well as two complete applications, in the \lmd{1-probing model} and \lmd{RR 1-probing model}. Verification is performed on five distinct CPUs (\cocoibex{}, \ibex{}, \cv{}, \cmthree{}, \cmfour{}). To our knowledge, this marks the first formal verification in the \lmd{RR 1-probing model} of a full first-order masked AES implementation running across several CPUs, with \framework{} completing verification in only 19 to 39 minutes, depending on the core. Finally, we compare our verification findings against eight real power measurements on two Arm CPUs. We observe that both original versions, flagged by \framework{} as leaky, do exhibit leakage in practice, whereas the hardened versions, certified as secure by our tool, show no practical leakage. This strongly supports the effectiveness of our approach for formal leakage verification.

All benchmarks---including hardware accelerators, masked programs, and CPU models (excluding the ARM cores)---are made available alongside \framework{}.

\bigskip


In the following, we first go through the background key verification notions and existing works (Section~\ref{sc:background}) before explaining our method and its formal definitions (Section~\ref{sc:method}). After a presentation of the implementation (Section~\ref{sc:implementation}), we present experimental results and discuss our approach (Section~\ref{sc:results}) before finally concluding on our work (Section~\ref{sc:conclusion}).

\section{Background on Verification of Masked Implementations}
\label{sc:background}

Various approaches have been proposed to assess the security of masked software implementations. Two main types of verification coexist: statistical approaches, which rely on statistical analysis applied to simulated or real traces, and formal approaches, which aim to verify the absence of leakage during program execution w.r.t. a given leakage model.

\subsection{Verification Based on Statistical Analysis}

The statistical analysis based approaches have been studied in various works within the previous decade.

\tool{ELMO}~\cite{EPRINT:McCWhiOsw16} is an instruction level power simulator which can generate power traces for the ARM Cortex-M0 and Cortex-M4 processors. Instruction power consumption is estimated using linear regression on real measurements per instruction. \tool{ELMO*}~\cite{NDSS:SSBRWY21} is an enhanced version of \tool{ELMO} which accounts notably for non-consecutive instructions and memory accesses. This simulator can then be used for fast verification using statistical metrics such as the $t$-test or the SNR. It is integrated in \tool{ROSITA}~\cite{NDSS:SSBRWY21} and \tool{ROSITA++}~\cite{CCS:SCSWBY21}, two code rewrite engines able to secure respectively first-order masked and higher-order masked implementations by exploiting \tool{ELMO*} outputs.

\tool{MAPS}~\cite{COSADE:CorGroDin18} is a micro-architectural power simulator for the Cortex-M3. The Cortex-M3 model includes only the registers involved in the manipulation of data, identified in the HDL description. \tool{MAPS} simulates ARM binaries to help identify leakages, in particular those introduced by the pipeline registers.

\tool{\prolead{}}~\cite{TCHES:MulMor22} is a verifier based on simulations and statistical hypothesis tests. By simulating the netlist of a circuit with numerous input vectors while placing probes at key locations, \prolead{} computes statistical independence with the G-test and is able to verify circuits in the \lmd{RR d-probing} model. As it is based on concrete simulations, the certainty of the results grows with the number of tested input vectors. \prolead{} can be used on large stateful (i.e. in which inputs are not fixed over time) circuits.

\tool{Prolead\_sw}~\cite{TCHES:ZeiMulMor23} extends \prolead{} to support the verification of masked software by simulating the execution of binaries running on a generic CPU model. It uses a CPU-independent abstract leakage model which encompasses various leakage sources reported in state-of-the-art works. It currently only supports ARM binaries.

\subsection{Formal Verification Methods}

Formal verification approaches, to prove whether a masked hardware or software implementation leaks within a defined leakage model, are usually split into two steps.

The first step consists in building a model of the target circuit or the target CPU running the masked software, representing all possible leakage sources in the chosen leakage model. Any input of the circuit or the software must be typed or labeled as \textit{public} for unsensitive data, and either \textit{masks} and \textit{secrets} or \textit{shares} for sensitive data.
From this model, a symbolic intermediate representation of all the data, sensitive or not, propagating through the circuit (or CPU) is extracted.

The second step is the application of a verification method 
to prove that the extracted symbolic intermediate representation of propagating data cannot leak information about the secrets in a given leakage model for all possible inputs. We identified three main verification methods. 
\begin{enumerate*}[1) ]
    \item The \emph{substitution}-based methods 
    rely on symbolic expressions to represent the intermediate values. They perform successive replacements of masked sub-expressions with a bijective mask, and aim at removing all secret variables from the expression, in which case the expression is statistically independent from all secrets. Introduced by~\cite{EC:BBDFGS15}, this method 
    is implemented in all the versions and variants of \tool{maskVerif}~\cite{ESORICS:BBCFGS19} as well as in \tool{VerifMSI}~\cite{verifmsi} and its previous version~\cite{leakageverif}.
    \item The \emph{inference}-based methods 
    also use symbolic expressions, along with types for describing different distributions of an expression, and rules to infer the distribution type of an expression from the types of its sub-expressions. Initially proposed by~\cite{jcenines} for verifying first order masked binary code, it has been extended with finer rules and to support more complex code structures~\cite{qmverif}.
    \item Finally, verification methods 
    based on \emph{constraints satisfiability} either express the absence of leakage in intermediate values with several constraints given to a SAT or SMT solver as in \coco{}~\cite{CocoAlma}; or represent intermediate values with specific structures such as ROBDD to efficiently determine their distribution, e.g. in the tools \tool{SILVER}~\cite{AC:KniSasMor20} and \prover{}~\cite{prover}.
\end{enumerate*} 

All mentioned formal verification methods 
except those based on ROBDD may have false positive. To discriminate between true and false positives, one has to determine the statistical independence of an expression w.r.t the secret values as presented in~\cite{enumhassan}, Figure 1. A resort consists in enumerating all possible values for symbolic inputs to compute this statistical independence. Note however, that this is not possible for large symbolic input domains.


In Section~\ref{sec:formal-verif-cpu}, we review existing work on the formal verification of masked software that takes into account micro-architectural leakage sources of the target CPU. In the following section, we discuss the impact of verification granularity, as previously introduced.


\subsection{Probe Width in Probing Models}

\label{sc:probe-width-models}

Since the earliest work on provable masking~\cite{C:IshSahWag03}, pen-and-paper verifications of generic masking schemes~\cite{C:IshSahWag03,CHES:RivPro10} have been considering a \lmd{d-probing} model, in which probes have an implicit width defined by the analysed description. In particular, it is 1-bit for masked circuit expressed at gate level~\cite{C:IshSahWag03},  
and the width of the intermediate computations for masked algorithm, e.g. $n$-bit for the masked multiplication over $\mathbb{F}_{2^n}$ algorithm or 8-bit for the masked AES~\cite{CHES:RivPro10}.

Automated verification methods presented in the previous section mainly target one kind of implementation, either hardware or software, at a specific level such as RTL or gate level for hardware, and such as assembly or pseudo-code level for software. In these verification methods, the probe width is always implicit: 1-bit for the hardware~\cite{ESORICS:BBCFGS19,AC:KniSasMor20,prover} and 8-bit for the software~\cite{jcenines,qmverif}.

When it comes to the verification of a program running on a CPU while considering its micro-architecture, one question arises: what should be the width of the probes? This also questions the probe width to consider when verifying hardware.

The probe width defines which signals or variables can be seen ``all at once''.
The issue regarding masked software is that the masking scheme support width can be different from the one carrying the information in hardware or at a lower level. For example, \tool{int8\_t} variables in \tool{C} will be manipulated on 32-bits registers and wires on a 32-bit ISA, possibly leading to more than one variable on a wire or register, depending on the compiler optimisations. 
Moreover, the probe width can seem totally arbitrary from a hardware perspective as in the final circuit wire or register bits are not necessarily close to each other. 
However, the probe width has a real impact on the security guarantee delivered by the verification. Considering only 1-bit signals implies that the eight bits of a byte wire can all be correctly masked with the same random bit; e.g. a masked design using only the two masks 0x00 and 0xFF would be proven secure if only 1-bit signals are verified, but would obviously not be proven secure at byte level, i.e. when considering 8-bit signals. Therefore, the verified security property is not the same whether considering 1-bit or wider signals, and wider signals verification brings a stronger security guarantee.

First, for clarity and reproducibility, we argue that the probe width should be made explicit in any formal verification setting. Second, from a security perspective — particularly in software verification based on hardware descriptions or at the ISA level — we advocate taking into account the width of the hardware elements that carry or store sensitive data.
In the context of hardware verification, if the goal is to verify a masking scheme under the assumption that each bit of a wire leaks independently from the others, then verification can be performed on a per-bit basis. Conversely, to detect violations of this independence assumption or to support masking schemes where all bits of a wire are assumed to leak jointly, verification must be carried out on the fully reconstructed wire.
To explicitly capture the probe width and distinguish between these two verification paradigms, we propose to refine the \lmd{d-probing} leakage model as either the \lmd{d-bit-probing} model or a support-wise leakage model: the \lmd{d-sw-probing} model. In the following, the \lmd{d-sw-probing} model is always considered with stability, as in the \lmd{RR d-probing} leakage model.

A limitation of the \lmd{d-sw-probing} model is that it depends on the hardware design choices, e.g. a processor could be described in HDL using only 1-bit signals. However, we expect that most of the time, the signal definitions in HDL are relevant w.r.t. the masking scheme and the desired leakage model. 

\subsection{Software Verification while Considering CPU Micro-Architecture}
\label{sec:formal-verif-cpu}

In this section, we present the existing approaches and associated tools for the formal verification of masked software running on CPUs, while considering the micro-architecture of the CPU. These approaches are aggregated in Table~\ref{tb:comparaison-soa}.

\begin{table}[ht!]
\small
\def\arraystretch{1.15}
\centering
\caption{Comparison of formal verification methods of software running on CPU.}
\label{tb:comparaison-soa}
\begin{tabular}{l|c|c|c|c}
\multirow{2}{*}{Feature/Tool} & \multirow{2}{*}{Armistice} & \multirow{2}{*}{Coco} & Closing & \multirow{2}{*}{\framework{}} \\
&&& the gap &\\ \hline
Multi-bit signals & \cmark & \xmark & \cmark & \cmark \\ \hline
Probing model & \lmd{(g=0,t=1) 1-sw} & \lmd{RR d-bit} & \lmd{RR d-bit} & \lmd{RR 1-sw} \\ \hline
Cryptographic & \multirow{2}{*}{\xmark} & \multirow{2}{*}{\cmark} & \multirow{2}{*}{\xmark} & \multirow{2}{*}{\cmark} \\
accelerators &&&& \\ \hline
Automatic & \multirow{2}{*}{\xmark} & \multirow{2}{*}{\cmark} & \multirow{2}{*}{\xmark\footnotemark} & \multirow{2}{*}{\cmark} \\
from HDL &&&& \\ \hline
\multirow{3}{*}{Usable CPU(s)} & Generic & Coco-Ibex, & \multirow{3}{*}{Ibex} & Cortex-M3, Cortex-M4 \\
 & CPU & Ibex, && Ibex, Coco-Ibex, \\
 & model & SweRV && CV32E40P \\ \hline
Memory model & \cmark & \cmark & \xmark & \cmark \\
\end{tabular}
\end{table}
\footnotetext{The contract redaction is a manual CPU-dependent task that must be performed before being verified.}

\tool{Armistice} is devoted to the simulation of a model of the Cortex-M3 datapath in order to verify masked software implementations running on it to cover leakages stemming from micro-architectural features. The model have been manually created from the Verilog description of the Cortex-M3, which is a highly non-portable and error-prone task, advocating for an automated extraction as we propose. Besides, \tool{Armistice} only supports value-based and transition-based leakage models. Still, \tool{Armistice} handles multi-bit signals as defined by the manually modelled Cortex-M3 and allows for support-wise verification.



\coco{}~\cite{CocoAlma} is a SAT-based formal verifier that can be used on programs running on CPU as well as hardware accelerators. It has been used on the RISC-V Ibex and SweRV CPUs~\cite{USENIX:GHPMB21,AC:GigPriMan21}. From an input labelling and a HDL description, it drives a simulation to extract from the simulation trace, and a model of the hardware target, 
the intermediate data to verify while considering transitions, glitches as well as stability. \coco{} only performs bit-probing verification.

\tool{Closing the gap}~\cite{closingthegap} is a contract-based formal method only for masked software running a CPU. Contracts express, at the ISA level, all the potential leakages. They are first validated against the HDL CPU description to ensure they capture all the potential leakages. Subsequently, masked programs are verified using symbolic simulation and the contracts. The main advantage of this approach is that the verification of the hardware against the contract, while being computation-intensive, is only done once. Afterward, the verification of the program is only performed w.r.t. the contract instead of a CPU model, thereby alleviating the burden of the verification. This approach requires extensive knowledge on the CPU to manually establish the contract. Currently, only the \ibex{} CPU is supported, and memories are not accounted for. The method claims to handle "word-wise" verification but does not describe how the bits are gathered into words, and the authors only justify word granularity for performance reasons. Therefore, verifications are not performed considering a support-wise probing model.

\tool{\framework{}} is the implementation of a new method combining a \emph{mixed-domain simulation} for the modelling of the circuit, and a substitution-based approach for the verification step. Its versatility allows for verification in variations of the \lmd{RR 1-sw-probing} model of both masked hardware and masked software running on several CPUs. The HDL description of the circuits and micro-architectures are directly exploited to generate intermediate representations of the models without particular knowledge on the verified target. It handles multi-bit signals as defined in Section~\ref{sc:probe-width-models} and accounts for the micro-architectural leakage sources in the memory subsystem.


\section{Method}
\label{sc:method}

\subsection{Overview}

Traditional HDL simulation computes values in the Boolean domain for all signals at each cycle. We refer to mixed-domain simulation as a simulation extension, by enriching the signal with values in other domains. This allows for the verification of secret leakage absence in a circuit, in presence of transitions and glitches. The additional signal values include:
\begin{itemize}\setlength\itemsep{0em}
     \item Symbolic expressions that describe wire values with respect to sensitive information such as shares, secrets, and masks.
     \item Sets of symbolic expressions, referred to as \lss{}, which represent all possible observable values when considering glitches.
     \item Stability information, for the eventual refinement of \lss{}, needed for the \lmd{RR 1-sw-probing} model.
\end{itemize}
Traditional concrete values are also computed, in particular for validation, allowing to check the correctness of the generated symbolic expressions. The simulation process thus occurs across four domains: the concrete domain, the symbolic domain, the \ls{} domain, and the stability domain. 

The flow of the method is depicted in Figure~\ref{fig:method}. The first input is a circuit which is either a CPU or a cryptographic accelerator, given in a Hardware Description Language (HDL). In the case of a processor, a binary program must also be provided. The other inputs are a leakage model, up to the \lmd{RR 1-sw-probing} model, stimuli needed to simulate the circuit such as clock or reset signals for all cycles and labels that maps the symbolic variables to their type, being either \textit{share}, \textit{secret}, \textit{mask} or \textit{public}.

\begin{figure}[ht]
    \centering
    \begin{tikzpicture}[node distance=0.15cm and 1cm]
    \tikzstyle{input} =   [align=center, text width=3cm]
    \tikzstyle{output} =  [text width=1.75cm]
    \tikzstyle{process} = [draw=black, align=center, inner sep=2mm, rounded corners=0.25cm, fill=Paired-8]
    \tikzstyle{arr} =     [arrows = -{Latex[width=2mm]}]
    \tikzstyle{pointer} = [draw=black, circle, font=\small, inner sep=0.5mm]

    \node (var) [input]                     {Labels};
    \node (lm)  [input, above=of var]       {Leakage Model};
    \node (sti) [input, below=of var]       {Input Stimuli};
    \node (bin) [input, below=of sti]       {Program};

    \node (rtl) [input, below=0.675cm of bin] {HDL Description};

    \node (gen) [process, right=of rtl]       {Model\\ Generator};
    \node (sim) [process, above=0.5cm of gen, inner ysep=0.34cm] {Mixed\\ Domain\\ Simulator};
    \node (manager) [process, below right=of sim.north east]   {Verification\\ Manager};
    \node (verif) [process, fill=Paired-7, right=of manager] {External\\ Verifier};

    \node (leak) [below=of verif] {Leaking wires};

    \node[draw=ccpuspec, dashed, inner sep=0.5mm, fit=(bin)](cpubox) {};
    \node[draw=cmethod, line width=0.8pt, inner sep=2.25mm, fit=(gen) (sim) (manager)](methodbox) {};

    \draw [arr] ([xshift=-0.25cm]rtl.east) -- (gen);
    \draw [arr] (gen.north) -- (sim.south);
    
    \draw [arr] ([xshift=-0.25cm]var.east) -- (var-|sim.west);
    \draw [arr] ([xshift=-0.25cm]sti.east) -- (sti-|sim.west);
    \draw [arr] ([xshift=-0.25cm]bin.east) -- (bin-|sim.west);

    \draw [arr] ([xshift=-0.25cm]lm.east) -| (manager.north);
    \draw [arr] ([yshift=0.25cm]sim.east|-manager.west) -- ([yshift=0.25cm]manager.west);
    \draw [arr] ([yshift=-0.25cm]manager.west) -- ([yshift=-0.25cm]sim.east|-manager.west);

    \draw [arr] ([yshift=0.25cm]manager.east) -- ([yshift=0.25cm]verif.west);
    \draw [arr] ([yshift=-0.25cm]verif.west) -- ([yshift=-0.25cm]manager.east);

    \draw [arr] (manager.south) |- (leak);

    \matrix [draw=black, above left] at ([xshift=0.15cm]current bounding box.south east) {
        \node [line width=0.8pt, inner sep=1.75mm, draw=cmethod, label=right:\framework{}] {}; \\
        \node [dashed, inner sep=1.75mm, draw=ccpuspec, label=right:CPU specific] {}; \\
    };

    \node[pointer, right=0.1cm] (num1) at ($(gen.north)!0.5!(sim.south)$) {1};
    \node[pointer, above=0.1cm] (num2) at ([yshift=0.25cm]$(sim.east|-manager.west)!0.5!(manager.west)$) {2};
    \node[pointer, above=0.1cm] (num3) at ([yshift=0.25cm]$(manager.east)!0.5!(verif.west)$) {3};
    \end{tikzpicture}
    \caption{Overview of \framework.}
    \label{fig:method}
\end{figure}

The flow begins with the generation, from the HDL description, of a model (step 1) that can be simulated while exposing all intermediate values needed for verification. This model is subsequently simulated (step 2) using user-provided inputs stimuli and labels. After each simulation cycle, the generated intermediate values are used by the verification manager which handles the interface with the external verifier, providing it with the relevant combinations of elements to verify in the specified leakage model (step 3). Reported leaks are output with the simulation cycle number, their description (i.e., which symbolic expression or \lss{}), their location (HDL source file, line number and wire name). This precise information can greatly help the designer understand and fix the reported secret leakages. The simulation (step 2) and verification (step 3) sequence iterates until the sequence of input stimuli have been entirely processed.

In the remainder of this section, we formally define the circuit model we consider, the different domains and the simulation in each domain.

\subsection{Circuit Modelling}

Hardware accelerators and CPUs running software are sequential circuits modelled in HDL and whose behaviour is modelled as functional Mealy machine. CPU targets generally embed data and instruction memories, which are here considered as part of the circuit.

\begin{definition}[Circuit]\label{def:circuit}
A circuit $\mathcal{C}$ is described with the tuple $\langle W, G, init \rangle$ where $G$ is the set of gates, which are either combinatorial ($\in C$) or registers ($\in R$) and $init$ is the vector of initial values for $R$. $W$ is the set of wires and contains $I$, the set of primary inputs, $O$, the set of primary outputs and other internal wires ($\in G \times G$). Any $w \in W$ has a width denoted $|w|$.
\end{definition}

\begin{definition}[Combinatorial gate]\label{def:comb-gate}
A combinatorial gate $g \in C$ is defined by the tuple $\langle f_g, W^{in}_{g}, w^{out}_{g}\rangle$ where $f_g$ is its functionality, $W^{in}_{g}$ is the set of its input wires and $w^{out}_{g}$ is its unique output wire. It can have multiple inputs but has a single output. 
\end{definition}

\begin{definition}[Sequential gate]\label{def:seq-gate}
A sequential gate a.k.a register, is a gate $r$ defined by the tuple $\langle w^{in}_{r}, w^{out}_{r}\rangle$ where $w^{in}_{r}$ and $w^{out}_{r}$ are its unique input and output wires. Sequential gates encompass memory elements.
\end{definition}


\begin{definition}[Functional Mealy machine]
\label{def:func-mealy}
The functional Mealy machine $\mathcal{M(C)}$ associated with a circuit $\mathcal{C}$ is described with the tuple  $\langle\Sigma_I, \Sigma_O, \Sigma_R, \delta, \lambda, \alpha_R^0\rangle$ with $\Sigma_I$, $\Sigma_O$, $\Sigma_R$ being respectively the sets of all valuations on $I$, $O$ and output wires of elements in $R$, this latest also denoted as the set of states.
\end{definition}

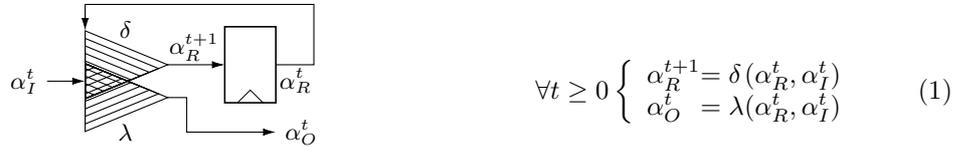
\begin{figure}[H]
    \centering
        \tikzstyle{register} = [flipflop, flipflop def={cd=1, t2={\hphantom}, t5={\hphantom}, clock wedge size=0.5}, scale=0.4, external pins width=0]
        \tikzstyle{logic} = [isosceles triangle, draw=black, inner sep=0, outer sep=0]
        \tikzstyle{ior} = [text centered, font={\small}]
        \tikzstyle{arr_automata} = [arrows = -{Latex[width=1mm]}]
        \begin{tikzpicture}[remember picture, node distance=0.5cm]
            \node (input) [ior] {$\alpha^t_I$};
            \node (TG) [ior, right=of input, minimum width=0cm, inner sep=0, outer sep=0] {};
            \foreach \i in {1,...,9}{
                \node (T\i) [logic, draw=black, above=-0.25cm of TG, anchor=right corner, minimum width=\i * 0.1cm] {};
            }
            \node (T) [ior, text=black, above=0cm of T9.north east] {$\delta$};
            \foreach \i in {1,...,9}{
                \node (G\i) [logic, draw=black, below=-0.25cm of TG, anchor=left corner, minimum width=\i * 0.1cm] {};
            }
            \node (G) [ior, text=black, below=0cm of G9.south east] {$\lambda$};
    
            \node (reg) [register, right=0.75cm of T9, line width=0.3pt] {};
            \node (state) [ior, below right=0.05cm and 0.05cm of reg.pin 5, minimum width=0cm, inner sep=0, outer sep=0] {$\alpha^t_R$};
            \node (stateafter) [ior, above left=0.05cm and 0.75cm of reg.pin 5, minimum width=0cm, inner sep=0, outer sep=0] {$\alpha^{t+1}_R$};
    
            \node (output) [ior, below right=0.1cm and 0cm of reg] {$\alpha^t_O$};
    
            \draw [arr_automata] (input) -- (TG);
            \draw [arr_automata] (T9) -- (reg.pin 2);
            \draw [arr_automata, arrows = -{Latex[width=1mm, reversed]}] (reg.east) -- ([xshift=0.5cm]reg.east) |- ([xshift=0.5cm, yshift=0.2cm]reg.north east) |- ([yshift=0.35cm]T9.left corner) |- ([yshift=0.17cm]T9.left corner);
            \draw [arr_automata] (G9.east) -- ([xshift=0.25cm]G9.east) |- (output);
        \end{tikzpicture}
    \caption{Functional Mealy machine.}
    \label{fig:mealy-machine}
\end{figure}

In Definition~\ref{def:func-mealy}, $\alpha_R^0 \in \Sigma_R$ is the initial state of $\mathcal{M(C)}$, it corresponds to the $init$ vector of $\mathcal{C}$.
More generally, we denote $\alpha^t \in \Sigma_W$, the valuation of all wires in $\mathcal{C}$ at cycle $t$, and $\alpha^t_X$ the restriction of $\alpha^t$ to the wire or set of wires $X$. In particular, $\alpha^t_R$ is called the state of the circuit at cycle $t$; $\alpha^t_I$ and  $\alpha^t_O$ are the input and output valuations at cycle $t$. A representation of a functional Mealy machine is given in Figure~\ref{fig:mealy-machine}.

In Definition~\ref{def:func-mealy}, $\delta$: $\Sigma_I \times \Sigma_R \to \Sigma_R$ is the transition function and $\lambda$: $\Sigma_I \times \Sigma_R \to \Sigma_O$ is the output function. Their application computes respectively the state for the next cycle and the current output valuations from the current input and current state, as given by Equation~\ref{eq:evolution-func-mealy}.
\begin{multicols}{2}
\columnbreak
\begin{equation}
\forall t \ge 0 \left\lbrace
\begin{array}{l@{}l@{}l}
    \alpha_R^{t+1} & = \delta  & (\alpha^{t}_R, \alpha^{t}_I) \\
    \alpha^{t}_O & = \lambda & (\alpha^{t}_R, \alpha^{t}_I) \\
\end{array}
\right.
\label{eq:evolution-func-mealy}
\end{equation}
\end{multicols}

Note that $\delta$ and $\lambda$ are invariant in time as the structure of the circuit is fixed. Moreover, as they compute a valuation for each element in $R$ and $O$ respectively, they both are vector-valued functions, composed of component functions, one for each element (of $R$ or $O$) to compute. 

The $\delta$ and $\lambda$ functions are built by applying the inductive traversal of the structure of the circuit $\mathcal{C}$ (Algorithm~\ref{alg:build-recursion}) on the input wires of each register (Equation~\ref{eq:build-delta}) and on each primary output (Equation~\ref{eq:build-lambda}) respectively.

\noindent\begin{tabularx}{\textwidth}{@{}XX@{}}
    \begin{equation}
        \delta = \left|
        \begin{array}{r@{}l}
             & \text{\textsc{build}}(w^{in}_{r_0}) \\
             & \vdots \\
             & \text{\textsc{build}}(w^{in}_{r_n}) \\
        \end{array}
        \right. r_0 \dots r_n \in R
        \label{eq:build-delta}
    \end{equation}
    &
    \begin{equation}
        \lambda = \left|
        \begin{array}{r@{}l}
             & \text{\textsc{build}}(w_0) \\
             & \vdots \\
             & \text{\textsc{build}}(w_n) \\
        \end{array}
        \right. w_0 \dots w_n \in O
        \label{eq:build-lambda}
    \end{equation}
\end{tabularx}

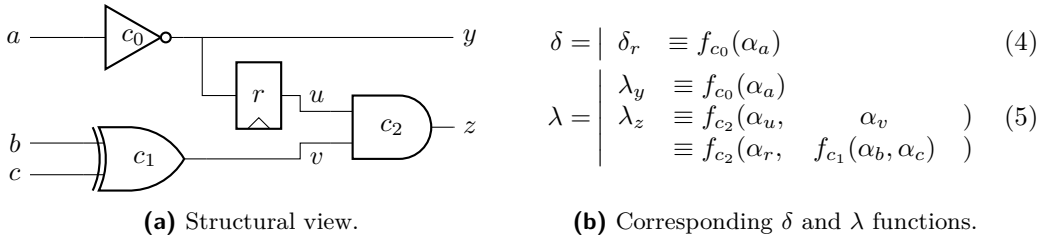
\begin{figure}[ht]
    \begin{subfigure}{0.49\textwidth}
        \tikzstyle{reg} = [flipflop, flipflop def={cd=1, t2={\hphantom}, t5={\hphantom}, clock wedge size=0.5}, scale=0.35, external pins width=0]
        \begin{tikzpicture}[node distance=0.35cm and 0.7cm]
            \node (i1) {$a$};

            \node[not port, right=of i1, anchor=in 1] (g0) {$c_0$};

            \node[xor port, below=1.4cm of g0.in 1, anchor=in 1] (g1) {$c_1$};

            \node[left=of g1.in 1] (i0) {$b$};
            \node[left=of g1.in 2] (i2) {$c$};
        
            \node[reg, below right=of g0, anchor=pin 2, font=\fontsize{30}{36}\selectfont] (r1) {$r$};
        
            \node[and port, xshift=2.5cm, anchor=center] (g2) at ($(r1.center)!0.5!(g1.center)$) {$c_2$};
        
            \node[anchor=south, yshift=-0.1cm] (w0) at ($(r1.pin 5)!0.5!(g2.bin 1)$) {$u$};
            \node[anchor=north] (w1) at (w0|-g2.bin 2) {$v$};
        
            \node[right=0cm of g2.out] (o1) {$z$};
            \node[] (o0) at (o1|-g0.out) {$y$};
        
            \draw (i1.east) -- (g0.in);
            \draw (i0.east) -- ([xshift=0.25cm]i0.east) |- (g1.in 1);
            \draw (i2.east) -- ([xshift=0.25cm]i2.east) |- (g1.in 2);
        
            \draw (r1.pin 5) -- ([xshift=0.25cm]r1.pin 5) |- (g2.in 1);
            \draw (g1.out) -- ([xshift=0.25cm]g1.out-|r1.pin 5) |- (g2.in 2);
        
            \draw (g0.out) -- ([xshift=0.25cm]g0.out) |- (r1.pin 2);
        
            \draw ([xshift=0.25cm]g0.out) -- (o0.west);
            \draw (g2.out) -- (o1.west);
        \end{tikzpicture}
        \caption{Structural view.}
        \label{fig:example-circuit-topo}
    \end{subfigure}
    \hfill
    \begin{subfigure}{0.50\textwidth}
        \begin{flalign}            
            \delta = &\left|
            \begin{array}{l@{}ll}
                & \delta_{r} &\equiv f_{c_0}(\alpha_a) \\
            \end{array}
            \right.\label{eq:example-delta} \\
            \lambda = &\left|
            \begin{array}{l@{}llcl@{}}
                & \lambda_{y} &\equiv f_{c_0}(\alpha_a) & & \\
                & \lambda_{z} &\equiv f_{c_2}(\alpha_u, & \alpha_v &) \\
                & &\equiv f_{c_2}(\alpha_r, & f_{c_1}(\alpha_b, \alpha_c) &) \\
            \end{array}
            \right.\label{eq:example-lambda}
        \end{flalign}
        \caption{Corresponding $\delta$ and $\lambda$ functions.}
        \label{fig:example-circuit-func}
    \end{subfigure}
    \caption{(a) A circuit example with  $I = \{a, b, c\}$, $O = \{y, z\}$ and internal the wires $\{u, v\}$. Gates sets are defined as $C = \{c_0, c_1, c_2\}$ and $R = \{r\}$. In this example, $f_{c_0}$ is a \textsc{not} functionality, $f_{c_1}$ a \textsc{xor} and $f_{c_2}$ a \textsc{and}. (b) The transition and output functions of the functional Mealy machine $\mathcal{M(C)}$.} 
    \label{fig:example-circuit}
\end{figure}

Figure~\ref{fig:example-circuit-func} provides the $\delta$ (Equation~\ref{eq:example-delta}) and $\lambda$ (Equation~\ref{eq:example-lambda}) functions of Mealy machine associated to the circuit example given in Figure~\ref{fig:example-circuit-topo} and obtained using Equations~\ref{eq:build-delta} and~\ref{eq:build-lambda} respectively. This circuit contains both combinatorial and sequential gates.

\begin{algorithm}
\caption{Inductive traversal of the structure of the circuit $\mathcal{C}$ from a wire $w$.}
\label{alg:build-recursion}
\DontPrintSemicolon
\SetKwInOut{Input}{Input}
\SetKwInOut{Output}{Output}
\Input{A wire $w$, which is either a primary input or the output wire of a gate $g$}
\Output{The functionality associated with $w$ up to the inputs and registers}
\SetKwFunction{build}{\textsc{build}}
\build{$w$}\textsc{:}\\
\Indp
\lIf{$w \in I$} {\Return $w$}
\lElseIf{$g \in R$} {\Return $w^{out}_{g}$}
\lElse{\Return{$f_g(\dots, $\build{$w_i$}$, \dots)$\hfill $\triangleright{}$ \textnormal{With $w_i$, the wires in $W^{in}_{g}$}}}
\end{algorithm}

Starting from $\alpha_R^0 \in \Sigma_R$, the Mealy machine, driven by the input configuration sequence $\alpha^0_I \dots \alpha^{n-1}_I \dots$, generates, through successive applications of $\lambda$ and resp. $\delta$, a sequence of states $\alpha_R^1 \dots \alpha_R^n \dots$ and resp. a sequence of output configurations $\alpha^0_O \dots \alpha^{n-1}_O \dots$. These intertwined sequences are represented as follows:
$$
\alpha_R^0 \xrightarrow[\alpha^0_O]{\alpha^0_I} \alpha_R^1 \xrightarrow{} \dots \xrightarrow{} \alpha_R^{n-1} \xrightarrow[\alpha^{n-1}_O]{\alpha^{n-1}_I} \alpha_R^n \xrightarrow[\dots]{\dots}\dots
$$

\subsection{Mixed Domain Simulation}

One application of the $\delta$ and $\lambda$ functions associated to $\mathcal{C}$ represents the simulation of the circuit during one clock cycle, while the successive applications of these functions simulate $\mathcal{C}$ over several clock cycles.
The usual simulation of a circuit is a simulation in the \textit{concrete} domain. It consists in computing logical values on each wire.

Security properties evaluation is based on the analysis of interaction of all sensitive data during computation over time. We introduce symbolic simulation to capture the exact contribution of each sensitive data to all intermediate computations, at each spatial and temporal point. The mixed-domain simulation extends the concrete simulation by propagating symbolic expressions representing all calculus w.r.t. the sensitive data, leading to a simulation in the \textit{symbolic} domain. The set of values which can be observed when considering glitches is propagated in the \ls{} domain and the information on stable signals is made in the \textit{stability} domain, to eventually refine the \ls{} computation.

We formally define these domains to represent the propagated values, along with the behaviour of each gate type in each domain, using dedicated functionality functions.

\begin{enumerate}\setlength\itemsep{0em}
    \item \textbf{Concrete Domain}: $\mathcal{B}$, the set of Boolean vectors of all sizes, $\mathcal{B} = \bigcup_{i\in \mathbb{N}^+} \dconc^i$, with $\dconc = \{0, 1\}$.
    \item \textbf{Symbolic Domain}: $\mathcal{E}$, the set of symbolic expressions of all widths, $\mathcal{E} = \bigcup_{i\in \mathbb{N}^+} \dsymb^i$, with $\dsymb^i$ the set of all expressions $e$ of width $i$ defined as follows:
\end{enumerate}
\begin{equation}
e \in \dsymb^i \Leftrightarrow e:= \left\lbrace
\begin{array}{r@{}l@{\hspace{0.23cm}}l@{}}
    & \token{CST}(x) & \text{a constant } x \in \dconc^i \\
    & \token{SYMB}(y) & y \text{, a symbolic variable of width } i \\ 
    & \token{OP\_*}(\dots, t_j, \dots) & \token{OP\_*} \text{, an operator of output width } i \text{ and } t_j \in \mathcal{E} \\
\end{array}
\right.
\label{eq:symb-domain-def}
\end{equation}

Table~\ref{tb:verif-msi-tokens} gives the complete list of operators which can appear as symbolic expression constructors.
\begin{table}[H]
    \centering
    \begin{tabular}{l|l|l|l}
        \token{OP\_XOR} & \token{OP\_ADD} & \token{OP\_LSL} & \token{OP\_CONCAT}  \\
        \token{OP\_AND} & \token{OP\_MUL} & \token{OP\_LSR} & \token{OP\_EXTRACT} \\
        \token{OP\_OR}  & \token{OP\_POW} & \token{OP\_ASR} & \token{OP\_ZEXT}    \\
        \token{OP\_NOT} & \token{OP\_SUB} & \token{ARRAY}   & \token{OP\_SEXT}    \\
    \end{tabular}
    \caption{List of function operators in the symbolic expression construction.}
    \label{tb:verif-msi-tokens}
\end{table}

\begin{enumerate}[resume]\setlength\itemsep{0em}
    \item \textbf{Leak Set Domain}: $\mathcal{L}$, the set of vectors of all sizes whose components are sets of symbolic expressions $\mathcal{L} = \bigcup_{i\in \mathbb{N}^+} \dlset^i$, with $\dlset = \mathscr{P}(\mathcal{E})$.
    \item \textbf{Stability Domain}: $\mathcal{S}$, the set of Boolean vectors of all sizes representing stability, $\mathcal{S} = \bigcup_{i\in \mathbb{N}^+} \dstab^i$, with $\dstab = \{0, 1\}$.
\end{enumerate}

\begin{definition}[Valuation of a wire] The valuation of a wire $w$ of width $|w|$ at cycle $t$ is the tuple $\alpha_w^t=\langle \vconc[t]{w}, \vsymb[t]{w}, \vlset[t]{w}, \vstab[t]{w}\rangle \in \dconc^{|w|} \times \dsymb^{|w|} \times \dlset^{|w|} \times \dstab^{|w|}$, where $\vconc[t]{w}$ is the concrete value of the wire at cycle $t$, $\vsymb[t]{w}$ its symbolic expression, \vlset[t]{w} its \ls{} and \vstab[t]{w} its stability. 
\end{definition}

In the following, we note $w[i]$ the $i^{\text{th}}$ bit a wire $w$.
We also note \fconc{f_g}, \fsymb{f_g}, \flset{f_g} and \fstab{f_g} the functionality of the gate $g$ in each domain.
For sake of clarity, only some gate functionalities in each domain are described (Tables~\ref{tb:functionalities-implem-concsymb},~\ref{tb:functionalities-implem-stab} and~\ref{tb:functionalities-implem-lset}).

\begin{figure}[ht]
    \centering
    \tikzstyle{tuple} = [align=center, font=\footnotesize]
    \begin{tikzpicture}[node distance=0.35cm and 2.5cm]

        \node[and port] (g0) {};
        \node[xor port, below right=of g0.out] (g1) {};
        \node[or port, below left=of g1.west] (g2) {};
        
        \node (i0) [left=of g0.in 1] {};
        \node (i1) [left=of g0.in 2] {};
        \node (i2) [left=of g2.in 1] {};
        \node (i3) [left=of g2.in 2] {};
        \node (o0) [right=of g1.out] {};
        
        \draw (i0.east) -- node [tuple, midway, above] {$\langle\textcolor{cconc}{1}, \textcolor{csymb}{X}, \textcolor{clset}{\{X\}}, \textcolor{cstab}{0}\rangle$} (g0.in 1);
        \draw (i1.east) -- node [tuple, midway, below] {$\langle\textcolor{cconc}{0},\textcolor{csymb}{\token{CST}(0)},\textcolor{clset}{\varnothing}, \textcolor{cstab}{0}\rangle$} (g0.in 2);
        
        \draw (i2.east) -- node [tuple, midway, above] {$\langle\textcolor{cconc}{0}, \textcolor{csymb}{Y}, \textcolor{clset}{\{Y\}}, \textcolor{cstab}{0}\rangle$} (g2.in 1);
        \draw (i3.east) -- node [tuple, midway, below] {$\langle\textcolor{cconc}{1}, \textcolor{csymb}{Z}, \textcolor{clset}{\{Z\}}, \textcolor{cstab}{1}\rangle$} (g2.in 2);
    
        \draw (g0.out) |- node [tuple, near end, above] {$\langle\textcolor{cconc}{0},\textcolor{csymb}{\token{CST}(0)}, \textcolor{clset}{\{X\}}, \textcolor{cstab}{0}\rangle$} (g1.in 1);
        \draw (g2.out) |- node [tuple, near end, below] {$\langle\textcolor{cconc}{1},\textcolor{csymb}{\token{OP\_OR}(Y,Z)},$\\ $\textcolor{clset}{\{\token{OP\_OR}(Y,Z)\}}, \textcolor{cstab}{1}\rangle$} (g1.in 2);
        
        \draw (g1.out) -- node [tuple, midway, above] {$\langle\textcolor{cconc}{1},\textcolor{csymb}{\token{OP\_OR}(Y,Z)},$\\ $\textcolor{clset}{\{X, \token{OP\_OR}(Y,Z)\}}, \textcolor{cstab}{0}\rangle$} (o0.west);
    \end{tikzpicture}
    \caption{Circuit valuations with $X=\token{SYMB}(x)$, $Y=\token{SYMB}(y)$ and $Z=\token{SYMB}(z)$.}
    \label{fig:example-simple}
\end{figure}
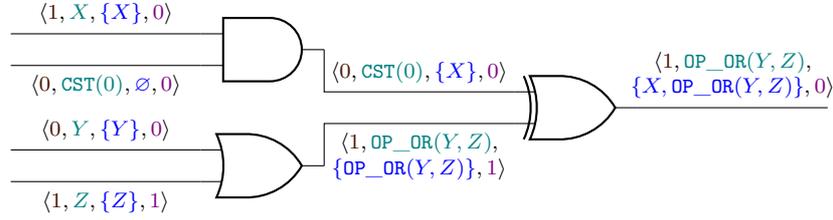

A simple circuit example is given in Figure~\ref{fig:example-simple} with valuations for each wire, in which symbolic expressions are simplified, e.g. $\token{OP\_AND}(\token{SYMB}(x), \token{CST}(0))$ becomes $\token{CST}(0)$.

For a wire $w$ at cycle $t$, the valuation $\alpha_w^t$ is computed with the application of an evaluation function for each domain, detailed in Sections~\ref{sc:concsymb-val},~\ref{sc:stability-val} and~\ref{sc:leakset-val}. Moreover, since $I$ represents the set of input wires, a valuation is provided for each domain by the input sequence for every instant, following Equation~\ref{eq:inputs-def}.
\begin{equation}
\alpha_{i \in I}^t = \left\langle \vconc[t]{i} \in \dconc^{|i|}, 
\vsymb[t]{i} = \left\lbrace
\begin{array}{@{}r@{}l@{}}
    & \token{CST}(\vconc[t]{i})\\
    & \token{SYMB}(name)\\
\end{array}
\right.,
\vlset[t]{i}[j] = \{\vsymb[t]{i}[j]\} \forall j<|i|,
\vstab[t]{i} = \vv{0} \right\rangle
\label{eq:inputs-def}
\end{equation}

\subsubsection{Concrete and Symbolic Valuations}

\label{sc:concsymb-val}

The evaluation functions of a wire $w$ for the concrete (Equation~\ref{eq:econc-def}) and symbolic (Equation~\ref{eq:symb-def}) domains for a cycle $t$, named \fconc{eval}($w$, $t$) and \fsymb{eval}($w$, $t$) respectively, are defined inductively until reaching inputs or registers along their cone of influence.
\begin{equation}
\fconc{eval}(w, t) := \left\lbrace
\begin{array}{r@{}ll}
    & \vconc[t]{w} & \text{if } w \in I \\
    & t=0 \text{ ? } \vconc[0]{w} : \vconc[t-1]{w^{in}_{r}} & \text{if } \exists r \in R \text{ s.t. } w = w^{out}_{r}\\
    & \fconc{f_g}(\dots, \fconc{eval}(w_i, t), \dots) & \text{else, } \exists g \text{ s.t. } w = w^{out}_{g}, \forall w_i \in W^{in}_g \\
\end{array}
\right.
\label{eq:econc-def}
\end{equation}

For both domains, if $w$ is the output wire of a register $r$, its valuation is the one of the input wire of $r$ at the previous cycle.
Otherwise, the valuation in both domains of a wire $w$ output of a gate $g$, is the application of the functionality function (Table~\ref{tb:functionalities-implem-concsymb}) \fconc{f_g} or \fsymb{f_g} to all its inputs projected in the domain. For the symbolic domain, if $w$ holds a symbolic constant \token{CST(x)}, \token{x} must correspond to the concrete valuation $\vconc[t]{w}$.
\begin{equation}
\fsymb{eval}(w, t) := \left\lbrace
\begin{array}{r@{}ll}
    & \vsymb[t]{w} & \text{if } w \in I \\
    & t=0 \text{ ? } \vsymb[0]{w} : \vsymb[t-1]{w^{in}_{r}} & \text{if } \exists r \in R \text{ s.t. } w = w^{out}_{r} \\
    & \fsymb{f_g}(\dots, \fsymb{eval}(w_i, t), \dots) & \text{else, } \exists g \text{ s.t. } w = w^{out}_{g}, \forall w_i \in W^{in}_g \\
\end{array}
\right.
\label{eq:symb-def}
\end{equation}
\begin{table}[ht]
    \def\arraystretch{1.15}
    \centering
    \begin{tabular}{|c|c|l|}
        \hline
        Domains & Inputs & Definition of \fconc{f_g} and \fsymb{f_g} \\ \hline
        $\dconc^n \times \dconc^n \to \dconc^n$ & $(\vconc{w1}, \vconc{w2})$ & $\fconc{f_g}(\dots) = \vconc{w1} \textsc{ and/or/xor } \vconc{w2}$ \\ \hline
        $\dsymb^n \times \dsymb^n \to \dsymb^n$ & $(\vsymb{w1}, \vsymb{w2})$ & $\fsymb{f_g}(\dots) = \token{OP\_\{AND/OR/XOR\}}\ (\vsymb{w1}, \vsymb{w2}$) \\ \hline
    \end{tabular}
    \caption{\textsc{and}, \textsc{or} and \textsc{xor} functionality computation in the $\dconc^n$ and $\dsymb^n$ domains.}
    \label{tb:functionalities-implem-concsymb}
\end{table}

\subsubsection{Stability Valuation}
\label{sc:stability-val}

The evaluation function \fstab{eval}($w$, $t$) for a cycle $t$ is bit-wise defined in Equation~\ref{eq:stab-def}, for readability reasons. It has the same termination conditions as for the concrete and symbolic domains.

By default, no information is available on the origin of the inputs signals. Hence, no assumptions should be made about their stability: they are always defaulted to unstable. 

When a wire is a register output, the stability of its bits is the bitwise syntactic equality of their current and previous symbolic expressions.

In most cases, the stability of the output of a combinatorial gate only depends on the stability of its inputs. However, particular input values for some gates allow to enforce stability using the symbolic valuation in conjunction with stability. For example, the stable symbolic input values \token{CST(0)} for \textsc{and} gates and \token{CST(1)} for \textsc{or} gates stabilise the output wire of these gates, regardless of the stability of the other input~\cite{proleadv3}. This requires that \fstab{f_g} takes as inputs the valuation of input wires in both the symbolic and the stability domains. The \textsc{and} and \textsc{or} functionalities in Table~\ref{tb:functionalities-implem-stab} illustrate such a possible finer behaviour while the \textsc{xor} and \textsc{not} functionalities only rely on their inputs stability valuations.

\begin{equation}
\fstab{eval}(w, t)[i]:=\left\lbrace
\begin{array}{r@{}ll}
    & \vstab[t]{w}[i] & \text{if } w \in I \\
    & t=0 \text{ ? } \vstab[0]{w}[i] : \vsymb[t]{w}[i]=\vsymb[t-1]{w}[i] & \text{if } \exists r \in R \text{ s.t. } w = w^{out}_{r} \\
    & \begin{array}{@{}l@{}l@{}}\fstab{f_g}(&\dots, \fstab{eval}(w_j, t), \dots, \\
    & \dots, \fsymb{eval}(w_j, t), \dots)[i]\end{array} & \begin{array}{@{}l@{}l@{}} &\text{else, } \exists g \text{ s.t. } w = w^{out}_{g}, \\ &\forall w_j \in W^{in}_g\end{array} \\
\end{array}
\right.
\label{eq:stab-def}
\end{equation}

\begin{table}[ht]
    \def\arraystretch{1.1}
    \centering
    \begin{tabular}{|@{\hspace{0.1cm}}c@{\hspace{0.1cm}}|c|c|@{\hspace{0.1cm}}l@{\hspace{0.1cm}}|}
    \hline
    \makecell{$f_g$} & \makecell{Domains} & Inputs & \makecell{Definition of \fstab{f_g}$[i]$} \\ \hline
    \multirow{3}{*}{\makecell{\textsc{and}}} & \multirow{3}{*}{\makecell{$\dstab^n \times \dstab^n \times \dsymb^n \times \dsymb^n$ \\ $\to \dstab^n$}} & 
    \multirow{3}{*}{$(\vstab{w1}, \vstab{w2}, \vsymb{w1}, \vsymb{w2})$} & 
    \multirow{3}{*}{\makecell[l]{
            \hspace{0.05cm} $(\vstab{w1}[i] \land \vstab{w2}[i])$ \\
            \hspace{0.05cm} $\lor~ (\vstab{w1}[i] \land (\vsymb{w1}[i] = \token{CST}(0)))$ \\
            \hspace{0.05cm} $\lor~ (\vstab{w2}[i] \land (\vsymb{w2}[i] = \token{CST}(0)))$
    }} \\
    &&& \\
    &&& \\ \hline
    \multirow{3}{*}{\makecell{\textsc{or}}} & \multirow{3}{*}{\makecell{$\dstab^n \times \dstab^n \times \dsymb^n \times \dsymb^n$ \\ $\to \dstab^n$}} & 
    \multirow{3}{*}{$(\vstab{w1}, \vstab{w2}, \vsymb{w1}, \vsymb{w2})$} & 
    \multirow{3}{*}{\makecell[l]{
            \hspace{0.05cm} $(\vstab{w1}[i] \lor \vstab{w2}[i])$ \\
            \hspace{0.05cm} $\lor~ (\vstab{w1}[i] \land (\vsymb{w1}[i] = \token{CST}(1)))$ \\
            \hspace{0.05cm} $\lor~ (\vstab{w2}[i] \land (\vsymb{w2}[i] = \token{CST}(1)))$
    }} \\
    &&& \\
    &&& \\ \hline
    \multirow{2}{*}{\makecell{\textsc{xor}}} & \multirow{2}{*}{\makecell{$\dstab^n \times \dstab^n \times \dsymb^n \times \dsymb^n$ \\ $\to \dstab^n$}} & 
    \multirow{2}{*}{$(\vstab{w1}, \vstab{w2}, \vsymb{w1}, \vsymb{w2})$} & 
    \multirow{2}{*}{\hspace{0.05cm} $\vstab{w1}[i] \land \vstab{w2}[i]$} \\
    &&& \\ \hline
    \makecell{\textsc{not}} & \makecell{$\dstab^n \times \dsymb^n \to \dstab^n$} & 
    $(\vstab{w}, \vsymb{w})$ & \hspace{0.05cm} $\vstab{w}[i]$ \\ \hline
    \end{tabular}
    \caption{Stability computation.}
    \label{tb:functionalities-implem-stab}
\end{table}

\subsubsection{Leakset Valuation}
\label{sc:leakset-val}

The evaluation function \flset{eval}($w$, $t$) is also bit-wise defined (Equation~\ref{eq:lset-def}) for readability reasons. \lss{} are vectors of symbolic expressions sets. For an input wire, the initial valuation is a vector of size $|w|$ containing for each of its elements, the singleton composed of the bit of the symbolic expression of the same rank. If the symbolic valuation is a \token{CST}, the empty set $\varnothing$ is used instead, this case is referred to as \emph{trivial} in the remainder of the paper. 

For registers, as both the expressions of the output wire at the current and the previous cycles can leak, the \ls{} for cycle $t$ is defined as the union of the sets of the same ranks. If a register output is stable, both expressions are identical, leading to a vector of singletons, analogously to the input \lss{} computation.

For a gate $g$, the \ls{} is computed on the inputs \lss{} as well as on its output stability and symbolic value. When the gate output is unstable, its valuation is a functionality-dependent combination of the inputs \lss{}. Whereas, when it is stable, its valuation is the vector of singletons containing the symbolic expression valuation bits (Table~\ref{tb:functionalities-implem-lset}).

As for inputs, during the evaluation of a register output wire or a combinatorial gate \ls{}, sets only composed of symbolic constants are always replaced by empty sets.

\begin{equation}
\flset{eval}(w, t)[i]:= \left\lbrace
\begin{array}{r@{}ll}
    & \{\vsymb[t]{w}[i]\}& \text{if } w \in I \\
    & t=0 \text{ ? } \vlset[0]{w}[i] :  \{\vsymb[t]{w}[i], \vsymb[t-1]{w}[i]\} & \text{if } \exists r \in R \text{ s.t. } w = w^{out}_{r} \\
    & \begin{array}{@{}l@{}l@{}}\flset{f_g}(&\dots, \flset{eval}(w_j, t), \dots, \\ &\fstab{eval}(w, t), \fsymb{eval}(w, t))[i]\end{array} & \begin{array}{@{}l@{}l@{}} &\text{else, } \exists g \text{ s.t. } w = w^{out}_{g}, \\ &\forall w_j \in W^{in}_g\end{array} \\
\end{array}
\right.
\label{eq:lset-def}
\end{equation}

\begin{table}[ht]
    \centering
    \def\arraystretch{1.25}
    \begin{tabular}{|c|c|l|}
        \hline
        \makecell{Domains} & Inputs & \makecell{Definition of \flset{f_g}[i]} \\ \hline
        \multirow{2}{*}{\makecell{$\dlset^n \times \dlset^n \times \dstab^n$ \\ $\times \dsymb^n \to \dlset^n$}} & \multirow{2}{*}{\makecell{$(\vlset{w1}, \vlset{w2},$\\ $ \vstab{w}, \vsymb{w})$}} & 
        \multirow{2}{*}{\makecell{$\flset{f_g}(\dots)[i] = \left\lbrace
            \begin{array}{l@{}l@{}l@{}}
            \vlset{w1}[i] \cup \vlset{w2}[i] & \text{ if } \neg\vstab{w}[i] \\
            \{\vsymb{w}[i]\} & \text{ otherwise} \\
            \end{array}
        \right.$}} \\
        && \\ \hline
    \end{tabular}
    \caption{\textsc{and}, \textsc{or} and \textsc{xor} functionality computation in the $\dlset^n$ domain.}
    \label{tb:functionalities-implem-lset}
\end{table}

The stability evaluation requires the expression valuation, while the \lss{} evaluation requires both the expression and stability valuation. For efficiency reasons, the computation of the four valuations should be performed in a specific order. We present such an order in Algorithm~\ref{alg:build-implem}, which performs a unique inductive traversal for a given wire at a given cycle, computing valuations in all the domains at once. By keeping in memory, during the cycle lifetime, all already computed valuations for internal wires, this traversal computes the state for the next cycle and the output for the current cycle.



\begin{algorithm}
\caption{Mixed-domain simulation of the valuation of a wire $w$ at cycle $t$.}
\label{alg:build-implem}
\DontPrintSemicolon
\SetKwInOut{Input}{Input}
\SetKwInOut{Output}{Output}
\Input{A wire $w$, which is either a primary input or the output of a gate $g$}
\Output{The valuation $\alpha_w^t$ of the wire $w$ at cycle $t$}
\SetKwFunction{eval}{\textsc{eval}}
\eval{$w$,$t$}\textsc{:}\\
\Indp
\lIf{$w \in I$} {\Return $\alpha_w^t$}
\lElseIf{$ g \in R$} {\Return $(t\ =\ 0\ ?\ \alpha_{w^{out}_{g}}^{0}\ :\ \alpha_{w^{in}_{g}}^{t-1})$}
\Else{
\tcc{With $w_{i1}\dots w_{in}$, the wires in $W^{in}_{g}$}
$\alpha_{w_{i1}}^t=\ $\eval{$w_{i1}, t$}\\
$\vdots$\\
$\alpha_{w_{in}}^t=\ $\eval{$w_{in}, t$}\\
$\textcolor{cconc}{cv}=\fconc{f_g}(\vconc[t]{w_{i1}}, \dots, \vconc[t]{w_{in}})$ \hfill $\triangleright{}$ \textnormal{Concrete value} \\
$\textcolor{csymb}{se}=\fsymb{f_g}(\vsymb[t]{w_{i1}}, \dots, \vsymb[t]{w_{in}})$ \hfill $\triangleright{}$ \textnormal{Symbolic expression} \\
$\textcolor{cstab}{st}=\fstab{f_g}(\vstab[t]{w_{i1}}, \dots, \vstab[t]{w_{in}}, \vsymb[t]{w_{i1}}, \dots, \vsymb[t]{w_{in}})$ \hfill $\triangleright{}$ \textnormal{Stability} \\
$\textcolor{clset}{ls}=\flset{f_g}(\vlset[t]{w_{i1}}, \dots, \vlset[t]{w_{in}}, \vsymb[t]{w}, \vstab[t]{w})$ \hfill $\triangleright{}$ \textnormal{LeakSet} \\
\Return{$\langle \textcolor{cconc}{cv}, \textcolor{csymb}{se}, \textcolor{clset}{ls}, \textcolor{cstab}{st}\rangle$}
}
\end{algorithm}

\subsection{Verification Management}
\label{sc:optimisations}

The verification is done separately from the simulation of the model. A \emph{verification manager} is responsible to extract simulated valuations in each domain and realise the appropriate verification in the given leakage model. The interface between the verification manager and the verifier is the function $\mathit{verif}$ which takes as input a set of expressions in $\mathcal{E}$ and outputs whether they leak altogether by ensuring that the joint distribution of the expressions values is independent from the secret values it contains. The verifier is not aware of the leakage model; hence, the verification manager builds an expression set that is consistent with the considered leakage model by using the valuations computed through the mixed-domain simulation. We explain how the expression set is built for each leakage model in the next section; we then explain the optimisation regarding the number of wires which must be verified at each cycle, depending on the leakage model.

\subsubsection{Expressions to Verify}

Table \ref{tb:expr-to-verify} gives the expression set built for a wire $w$ at a cycle $t$ for each value of \lmd{g} and \lmd{t} in the \lmd{(g,t) 1-sw-probing} leakage model.

\begin{table}[H]
\def\arraystretch{1.25}
\centering
\begin{tabular}{|c|l|}
\hline
\lmd{(g,t) 1-sw-probing} model & Set of expressions to verify \\ \hline
\lmd{(g=0, t=0)} & $\{\vsymb[t]{w}\}$\\ \hline
\lmd{(g=0, t=1)} & $\{\vsymb[t]{w}, \vsymb[t-1]{w}\}$ \\ \hline
\lmd{(g=1, t=0)} & $\bigcup_{i=0}^{|w|-1}\vlset[t]{w}[i]$ \\ \hline
\lmd{(g=1, t=1)} & $\{\vsymb[t-1]{w}\} \cup (\bigcup_{i=0}^{|w|-1}\vlset[t]{w}[i])$ \\ \hline
\end{tabular}
\caption{Expressions set built for a wire $w$ at cycle $t$ for each leakage model.}
\label{tb:expr-to-verify}
\end{table}

\noindent\textbf{Value leakage.} For the \lmd{(0,0) 1-sw-probing} leakage model, the singleton composed of the current expression of $w$, namely $\vsymb[t]{w}$, is sent to the verifier. 

\noindent\textbf{Transition leakage.} For the \lmd{(0,1) 1-sw-probing} model, $\{\vsymb[t]{w}, \vsymb[t-1]{w}\}$ the set containing the expressions from the current cycle and the previous one is sent to the verifier. This catches all possible combinations of both expressions, hence any possible transition. 

\noindent\textbf{Glitch leakage.} For the \lmd{(1,0) 1-sw-probing} model, the sets of expressions in the \ls{} are flattened in a set that is sent to the verifier. This set captures all the glitches at the wire level.

\noindent\textbf{Transition and glitch leakage.} For the \lmd{(1,1) 1-sw-probing} model, the expression set sent is the singleton containing $\vsymb[t-1]{w}$, the expression of $w$ at the cycle $t-1$, with all the expressions of the flattened set of $w$ at cycle $t$.

We can note that the verification of the expression built for the case with transitions or glitches encompasses the verification performed for the value leakage model. Similarly, the verification with both transitions and glitches encompasses the value leakage model, the transitions leakage model and the glitches leakage model.

\subsubsection{Wires to Verify}
\label{sc:wires-to-verify}

The straightforward verification strategy is to verify each wire at each cycle by sending to the verifier the expression set built according to the considered leakage model (cf. Table~\ref{tb:expr-to-verify}). In practice, many of the built expression sets are either empty or only composed of symbolic constants \token{CST(x)}, and thus are considered trivial and should not result in a verification request.

We explain in this section the optimisations regarding the number of verification requests that can be applied by the verification management, depending on the considered leakage model. Table~\ref{tb:wire-to-verify} gives for each leakage model the wires to verify at each cycle. 

\begin{table}[t]
\centering
\begin{tabular}{|l|p{8cm}|}
\hline
\lmd{(g,t) 1-sw-probing model} & Wires to verify \\ \hline
(0,0) & $w \in W$ (all the wires) \\ \hline
(0,1) & $w \in W$ (all the wires)\\ \hline
(1,0) & $w \in W^{in}_r$ for $r \in R$, $w \in O$, $w \in W^{split}$, \newline $w \in W^{in}_g$ with $\vstab[t]{w_g^{out}}$ $\neq \vv{0}$, \newline $w$ other input of multiplexer whose selector is stable \\ \hline
(1,1) & $w \in W$  (all the wires) \\ \hline
\end{tabular}
\caption{Wires $w$ to verify at cycle $t$ for each leakage model.}
\label{tb:wire-to-verify}
\end{table}

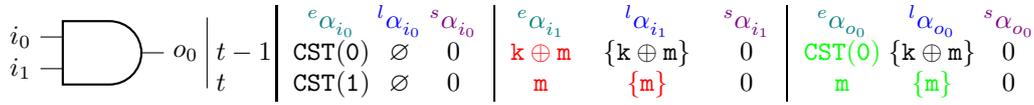
\begin{figure}[h!]
    \begin{subfigure}[c]{0.20\textwidth}
        \centering
        \tikzstyle{register} = [rectangle, minimum width=0.5cm, minimum height=1cm, text centered, draw=black]
        \tikzstyle{tuple} = [align=center, font=\footnotesize]
        \begin{tikzpicture}[node distance=0cm and 0cm]
            \node[and port] (g0) {};
            \node (i0) [left=of g0.in 1] {$i_0$};
            \node (i1) [left=of g0.in 2] {$i_1$};
            \node (o0) [right=of g0.out] {$o_0$};
            \draw[black] (current bounding box.south east) -- (current bounding box.north east);
        \end{tikzpicture}
    \end{subfigure}
    \hfill
    \begin{subfigure}{0.792\textwidth}          
        \begin{tabular}{@{}l@{\hspace{0.05cm}}|c@{\hspace{0.1cm}}c@{\hspace{0.1cm}}c|ccc|c@{\hspace{0.1cm}}c@{\hspace{0.1cm}}c@{}}
            & $\vsymb[]{i_0}$ & $\vlset[]{i_0}$ & $\vstab[]{i_0}$ & $\vsymb[]{i_1}$ & $\vlset[]{i_1}$ & $\vstab[]{i_1}$ & $\vsymb[]{o_0}$ & $\vlset[]{o_0}$ & $\vstab[]{o_0}$ \\
        $t-1$ & \token{CST(0)} & $\varnothing$ & 0 & \textcolor{red}{\token{k \oplus m}} & $\{\token{k \oplus m}\}$ & 0 & \textcolor{green}{\token{CST(0)}} & $\{\token{k \oplus m}\}$ & 0 \\
        $t$   & \token{CST(1)} & $\varnothing$ & 0 & \textcolor{red}{\token{m}} & \textcolor{red}{$\{\token{m}\}$} & 0 & \textcolor{green}{\token{m}} & \textcolor{green}{$\{\token{m}\}$} & 0 \\
        \end{tabular}
    \end{subfigure}
    \caption{Example illustrating the need to verify each wire in the \lmd{(0,1)} and \lmd{(1,1) 1-sw-probing} models. Input $i_1$ leaks \token{k} at cycle $t$ while output $o_0$ does not.}
    \label{fig:contre-exemple-allwires}
\end{figure}

\noindent\textbf{Value and transition leakage.} For the \lmd{(0,0)} and \lmd{(0,1) 1-sw-probing} leakage models, the verified expression set contains the current (and the previous, for transitions) expression of the wire. We cannot reduce the number of verified wires, as an expression computed by a gate $g$ can leak (e.g. $(k \oplus m) \oplus m$) while an expression of a gate $h$ taking $g$ as input does not ($m' \oplus ((k \oplus m) \oplus m)$). All the wires of the design must then be verified at each cycle. Figure~\ref{fig:contre-exemple-allwires} illustrates this requirement with an example in which the wire $i_1$ leaks when considering transitions whilst the result of the gate $o_0$ does not.

\begin{figure}[h!]
    \centering
    \tikzstyle{reg} = [flipflop, flipflop def={cd=1, t2={\hphantom}, t5={\hphantom}, clock wedge size=0.5}, scale=0.35, external pins width=0]
    \tikzstyle{tuple} = [align=center, font=\footnotesize]
    \begin{tikzpicture}[node distance=2cm and 2.1cm]
        \node (i) [] {};
        
        \node[reg, above right=-0.05cm and 7cm of i] (r0) {};
        \node[reg, below right=-0.05cm and 7cm of i] (r1) {};
            
        \draw (i.east) -- node [tuple, midway] (s0) {$/$} ([xshift=4.5cm]i.east);
        \draw ([xshift=4.5cm]i.east) -| ([xshift=-2.5cm]r0.pin 2) -- node [tuple, midway] (s1) {$/$} (r0.pin 2);
        \draw ([xshift=4.5cm]i.east) -| ([xshift=-2.5cm]r1.pin 2) -- node [tuple, midway] (s2) {$/$} (r1.pin 2);

        \node (ss0) [above=-0.2cm of s0] {$2$};
        \node (ss1) [above=-0.2cm of s1] {$1$};
        \node (ss2) [above=-0.2cm of s2] {$1$};
        \node (legi) [below=-0.2cm of s0] {\token{OP\_CONCAT(m, OP\_XOR(k, m) )}};
        \node (leg1) [below=-0.2cm of s1] {\token{m}};
        \node (leg2) [below=-0.2cm of s2] {\token{OP\_XOR(k, m)}};
    \end{tikzpicture}

    \caption{Example illustrating the need to verify wires which are split and partially used by a gate, with $\token{k}=\token{SYMB}(K)$ and $\token{m}=\token{SYMB}(M)$. The 2-bit wire leaks \token{k} while both 1-bit wires, each input of a register, do not.}
    \label{fig:exemple-weakened-word}
\end{figure}
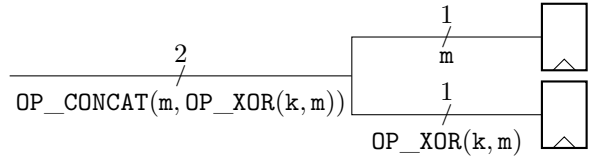



\noindent\textbf{Glitch leakage.} When only glitches are considered, not all wires need to be verified. For the \lmd{(1,0) 1-bit-probing} model without stability, \lss{} can only increase in size when traversing gates, and become maximal when reaching primary outputs of the design ($w \in O$) or registers ($w \in W^{in}_r$ for $r \in R$). However, when considering stability, this assumption does not hold anymore, because stability precisely allows the output \ls{} of a gate to be reduced. This is in particular the case for the inputs of gates whose outputs are stable (cf. Table~\ref{tb:functionalities-implem-lset}) or the non-selected inputs of \textsc{multiplexor} gates whose selectors are stable. Thus, these wires must be verified to account for the effect of stability. Finally, in the \lmd{(1,0) 1-sw-probing} model, \lss{} can also be reduced when a signal is split, as shown in Figure~\ref{fig:exemple-weakened-word}. Therefore, the set of wires that are split, which we denote $W^{split}$, must also be verified.

\noindent\textbf{Glitch and transition leakage.} For the \lmd{(1,1) 1-sw-probing} model, the built expressions include the expression of the wire at the previous cycle (cf. Table \ref{tb:expr-to-verify}). As in the value leakage model, we must verify all the wires. The example given in Figure~\ref{fig:contre-exemple-allwires} illustrates why this is also required for this leakage model.

\begin{figure}[h!]
    \begin{subfigure}[c]{0.20\textwidth}
        \centering
        \tikzstyle{register} = [rectangle, minimum width=0.5cm, minimum height=1cm, text centered, draw=black]
        \tikzstyle{tuple} = [align=center, font=\footnotesize]
        \begin{tikzpicture}[node distance=0cm and 0cm]
            \node[and port] (g0) {};
            \node (i0) [left=of g0.in 1] {$i_0$};
            \node (i1) [left=of g0.in 2] {$i_1$};
            \node (o0) [right=of g0.out] {$o_0$};
            \draw[black] (current bounding box.south east) -- (current bounding box.north east);
        \end{tikzpicture}
    \end{subfigure}
    \hfill
    \begin{subfigure}{0.792\textwidth}          
        \begin{tabular}{l|c@{\hspace{0.1cm}}c@{\hspace{0.1cm}}c|ccc|c@{\hspace{0.1cm}}c@{\hspace{0.1cm}}c}
            & $\vsymb[]{i_0}$ & $\vlset[]{i_0}$ & $\vstab[]{i_0}$ & $\vsymb[]{i_1}$ & $\vlset[]{i_1}$ & $\vstab[]{i_1}$ & $\vsymb[]{o_0}$ & $\vlset[]{o_0}$ & $\vstab[]{o_0}$ \\
        $t-1$ & \token{CST(0)} & $\varnothing$ & 1 & \token{k \oplus m} & \textcolor{red}{$\{\token{k \oplus m}\}$} & 0 & \token{CST(0)} & \textcolor{green}{$\varnothing$} & 1 \\
        $t$   & \token{CST(1)} & $\varnothing$ & 0 & \token{m} & \textcolor{red}{$\{\token{m}\}$} & 0 & \token{m} & \textcolor{green}{$\{\token{m}\}$} & 0 \\
        \end{tabular}
    \end{subfigure}
    \caption{Example showing the need to verify, at cycle $t$, the input wires of a stable gate at the cycle $t$ $-$ 1. In the \lmd{(1,1) 1-sw-probing} model, input $i_1$ leaks \token{k} at cycle $t$ while output $o_0$ does not.}
    \label{fig:exemple-past-stab}
\end{figure}
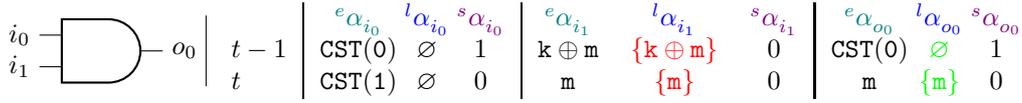

However, we can modify the expression set sent to the verifier (cf. Table \ref{tb:expr-to-verify}) to the one containing all the elements of flattened sets of $w$ at cycle $t$ and $t-1$ i.e. the expression: $(\bigcup_{i=0}^{|w|-1}\vlset[t-1]{w}[i]) \cup (\bigcup_{i=0}^{|w|-1}\vlset[t]{w}[i])$. This expression set does not include the expression of the wire at $t-1$ but its \ls{} at $t-1$, which is correct because it over-approximates the leakage of the expression. This enables to reduce the number of verification requests: we do not need to verify all wires, but only those described in Table~\ref{tb:over-approximation}. In particular, the input wires of stable gates at cycle $t-1$ must be verified at cycle $t$. Figure~\ref{fig:exemple-past-stab} presents a circuit that would not reveal secret leakage when applying the over-approximation if we did not consider the stability at $t-1$.

\begin{table}[h!]
\def\arraystretch{1.25}
\centering
\begin{tabular}{|l|p{9cm}|}
\hline
Expression to verify & $(\bigcup_{i=0}^{|w|-1}\vlset[t-1]{w}[i]) \cup (\bigcup_{i=0}^{|w|-1}\vlset[t]{w}[i])$. \\ \hline
Wires to verify & $w \in W^{in}_r$ for $r \in R$, $w \in O$, $w \in W^{split}$, \newline $w \in W^{in}_g$ with $\vstab[t]{w_g^{out}} \neq \langle 0\rangle$ or $\vstab[t-1]{w_g^{out}} \neq \langle 0\rangle$, \newline $w$ the non-selected inputs of a \textsc{mux} with stable selector at cycle $t$ or $t-1$ \\ \hline
\end{tabular}
\caption{Wires $w$ and associated expressions to verify at cycle $t$ for the \lmd{(1,1) 1-sw-probing} model with over-approximation.}
\label{tb:over-approximation}
\end{table}

Our approach considers \lmd{sw-probing} models, using the wires width specified in the HDL description of the design to analyse. However, one may want to consider 1-bit wires. In this case, using \lmd{(g,t) 1-bit-probing} models, all the reductions of the wires to verify presented above are still sound, but the verification of wires that are split is no more necessary. Verification in these more limited probing models does not enable the detection of violation of mask independence on a wire. Moreover, there are many more verification requests as all the bits of each wire to verify induce a verification request.

\section{Implementation}

\label{sc:implementation}

The section describes \framework{}, the framework that implements the mixed-simulation method proposed in Section~\ref{sc:method}. We first present the implementation of the verification flow, illustrated in the Figure~\ref{fig:implem-big-picture-new}, before explaining some of its key points.

\begin{figure}[ht]
    \centering
    \begin{tikzpicture}[remember picture, node distance=0.35cm and 0.75cm]
        \tikzstyle{input}     = [align=center, text width=1.75cm, font=\small]
        \tikzstyle{output}    = [text width=1.75cm, font=\small]
        \tikzstyle{arr}       = [arrows = -{Latex[width=2mm]}]
        \tikzstyle{darr}      = [arrows = {Latex[width=2mm]}-{Latex[width=2mm]}]
        \tikzstyle{element}   = [rectangle, font=\small, minimum width=1.75cm, text width=1.75cm, minimum height=1cm, text centered, draw=black]
        \tikzstyle{process}   = [shape=rounded rectangle, font=\small, minimum width=1.75cm, text width=1.68cm, minimum height=1cm, text centered, draw=black]
        \tikzstyle{overriden} = [draw=black, line width=1.2pt]

        \node (hdl) [input] {HDL (v/sv/vhdl)};
        \node (program) [input, above=of hdl] {Program (c/asm/o)}; 
        \node (label) [input, above=of program] {Labels};
        \node (stim) [input, above=of label] {Stimuli};
        \node (lmd) [input, above=of stim] {Leakage model};

        \node (init) [element, right=0.75cm of stim, overriden] {Init};
        \node (simu) [element, right=of init] {Simulation Driver};
        \node (manager) [process, right=of simu, overriden] {Manager};
        \node (verifmsi) [process, right=of manager] {VerifMSI++};
        \node (model) [element, below=0.55cm of simu, overriden] {Model};
        \node (runtime) [element, right=of model, overriden] {\tool{cxxrtl} runtime};

        \node (cxxrtl) [process, below=0.55cm of model, overriden] {\tool{cxxrtl} backend};
        \node (yosys) [process, left=of cxxrtl] {\tool{yosys}};

        \node[draw=ccpuspec, dashed, inner sep=0mm, fit=(program)](cpuspec) {};

        \draw [arr] (hdl.east|-yosys.west) -- (yosys.west);
        \draw [arr] (program) -| ([xshift=0.5cm]init.south);
        \draw [arr] (label) -| ([xshift=-0.5cm]init.south);
        \draw [arr] (stim) -- (init);
        \draw [arr] (lmd) -| (manager);

        \draw [arr] (yosys) -- (cxxrtl);
        \draw [arr] (cxxrtl) -- (model);
        \draw [arr] (runtime) -- (model);
        \draw [darr] (simu) -- (model);
        \draw [arr] (init) -- (simu);
        \draw [darr] (manager) -- (simu);
        \draw [darr] (verifmsi) -- (manager);

        \node[draw=cmethod, fill opacity=0.25, line width=0.8pt, inner sep=3mm, fit=(yosys) (cxxrtl) (manager)](methodbox) {};
        \node[draw=Paired-5, line width=0.8pt, inner sep=2.25mm, fit=(yosys) (cxxrtl)](modelgen) {};
        \node[draw=Paired-7, line width=0.8pt, inner sep=2.25mm, fit=(init) (simu)](modelsim) {};
        \node[draw=Paired-3, line width=0.8pt, inner sep=2.25mm, fit=(manager) (verifmsi)](modelverif) {};

        \matrix [draw=black, above left] at ([xshift=0.15cm]current bounding box.south east) {
            \node [inner sep=1.75mm, draw=cmethod, label=right:\framework{}] {}; \\
            \node [inner sep=1.75mm, draw=Paired-5, label=right:Model Gen.] {}; \\
            \node [inner sep=1.75mm, draw=Paired-7, label=right:Simulation] {}; \\
            \node [inner sep=1.75mm, draw=Paired-3, label=right:Verification] {}; \\
            \node [inner sep=1.75mm, overriden, label=right:Overriden] {}; \\
            \node [dashed, inner sep=1.75mm, draw=ccpuspec, label=right:CPU Only] {}; \\
        };
    \end{tikzpicture}
    \caption{Implementation of \framework.}
    \label{fig:implem-big-picture-new}
\end{figure}

\subsection{Implementation Flow}

\noindent\textbf{Model generation.} This first step is based on \tool{yosys}~\cite{Yosys} and a modified version of its concrete simulation backend, \tool{cxxrtl}.

\tool{Yosys} is provided with HDL source files in Verilog, SystemVerilog, or VHDL (via GHDL). It then performs a partial synthesis—applying all synthesis steps except technology mapping—first on the CPU core (if present), and then on the complete system-on-chip or hardware accelerator.

The \tool{cxxrtl} backend is subsequently invoked to generate a C++ model that can be simulated, which is compiled against a custom version of the \tool{cxxrtl} runtime embedding all domain evaluation functions. This compilation is performed once for all executions of the hardware or programs on CPU. Said programs are compiled independently and loaded at simulation time.

The gates handled by \framework{} are listed in Table~\ref{tb:lowlevel-func}. By operating at the level of high-level gates, we are able to verify circuits independently of the specific technological implementation of these gates, relying instead on the generic behavioural specifications, such as those provided as examples in Tables~\ref{tb:functionalities-implem-concsymb},\ref{tb:functionalities-implem-stab}, and~\ref{tb:functionalities-implem-lset}. While verifying a netlist mapped to a specific technology would also be feasible with \framework{}, the tool is not currently designed for that purpose.





\begin{table}[h]
    \centering
    \begin{tabular}{|c|c|c|c|c|c|}
    \hline
    Bitwise  & Comparison & Arithmetic & Shifts & Resize   & Miscellaneous \\ \hline
    bit\_not & ucmp       & add        & shl    & (r)trunc & is\_zero      \\
    bit\_and & scmp       & sub        & shr    & (r)zext  & is\_neg       \\
    bit\_or  & equal      & neg        & sshr   & sext     & mem\_write    \\
    bit\_xor & not\_equal & mul        &        & blit     & mem\_read     \\
             &            &            &        & repeat   & register      \\ \hline
    \end{tabular}
    \caption{Gates handled by \framework{}.}
    \label{tb:lowlevel-func}
\end{table}

\noindent\textbf{Simulation.} This second step is performed by feeding input stimuli to the model and triggering its evaluation. The CPU specific element presented in Figure~\ref{fig:implem-big-picture-new} is the program. \framework{} is able to handle any C, assembly or object program given by the user without recompiling the model by generating a plug-and-play memory loader including a bootloader for each program. Along with the program, the user provides the labels for the variables stored in memory or registers and can implement helper functions that are triggered at main function first and last instructions or at each cycle if needed. All programs are cross-compiled for the target architecture using the \tool{llvm} compilation infrastructure, ensuring a seamless integration in the \framework{}'s toolchain. The simulation driver computes for all wires, the valuations in the four domains.

\noindent\textbf{Verification.} This third step is performed by the verification manager, which extracts from the model simulation the valuations in all domains for all wires. Then, thanks to its memory of the previous cycle valuations, it performs the needed verifications for the configured 
leakage model. Simulation and verification are performed in an interleaved manner until the end of the program. Depending on the verification result and the configuration, the simulation can also terminate after the first secret leakage. \framework{} additionally provides the user with detailed information about the nature of the eventual secret leakages such as the simulation cycle number, their description (i.e., which symbolic expression, \lss{} or combination performed), their location (HDL source file, line number and wire name).

In order to build and manipulate its expressions, \framework{} relies on an external verifier called \tool{VerifMSI++}, which is an enhanced C++ version of \tool{VerifMSI}~\cite{verifmsi}, with bug fixes and more simplification rules. As a C++ library, it proves to be a good candidate for interfacing with \framework{}. \tool{VerifMSI++} allows the representation of complex symbolic expressions, their simplification, and implements an algorithm to determine if a given set of expressions statistically depends on the secret variables it contains, for a given security property.

\subsection{Simulator Modifications and Verification Manager}

Various adaptations of the concrete simulator \tool{cxxrtl} in both the backend, for the model generation, and the runtime --- now embedding the domain rules --- have been done.

\noindent\textbf{Model adaptation.} As we want to verify the output of each individual gate, \tool{cxxrtl} is configured to expose each wire of the netlist, i.e. it does not optimise parts of it by chaining gates in the simulation, but stores each gate output in the simulation state. For the same reasons, each implicit gates such as registers and multiplexers are made explicit in the generated model.

\noindent\textbf{Overloaded memory accesses.} Memory reads and writes are overloaded in such a way that, when specified, a user-defined function which takes as parameter the requested memory index wire, can be called. This enables the mixed-simulation of tabulated ciphers, which relies on a masking translation property on the symbolic memory reads (See \tool{AES-Herbst} example in Section~\ref{sc:results_sw}).
 
\noindent\textbf{Restrictions on circuits.} We restrict the circuits handled by \framework{} to those that can be represented as Mealy machines. They must also be exempt of combinatorial feedback loops, so that there exists an evaluation scheduling which enables the evaluation of each gate once to fully determine the outputs and next state. We only consider circuits with a single clock domain and in which the synchronous gates are all sensitive to the same clock edge. Besides, \tool{cxxrtl} has the following limitation: it considers that using a given bit of a signal to compute another bit of the same signal without crossing a register is a combinatorial feedback loop. For circuits exhibiting this behaviour, identifying these wires and splitting them is a required additional step to meet our assumptions. It is almost fully automated in our modified backend of \tool{cxxrtl} but an additional pass may be needed for certain circuits. Wire splitting applied during the partial synthesis pass is done with a unique name generator that allows for the verification manager to recombine these wires for the support-aware verification. The verification manager is also made aware of the structure of the circuit via a file generated by \tool{yosys} in addition to the model. This structural information is needed for the manager to detect which wires are a non-selected input of a multiplexer, as well as which wires are partially used (Figure~\ref{fig:exemple-weakened-word}).

\noindent\textbf{Interface for extracting valuations.} The verification manager, as described in Section~\ref{sc:wires-to-verify}, is in charge of extracting the valuations in the four domains from the model, for all internal wires. 
To this end, an interface allows the manager to query this information as well as to trigger a garbage collector for cleaning temporal \lss{}.

\noindent\textbf{Expression cache.} To enhance efficiency, a caching mechanism is implemented to verify identical expression sets only once. This optimisation significantly improves the implementation’s performance.

\section{Experimental Results}

\label{sc:results}

In order to prove the interest of the approach, we experimentally show on several benchmarks that \framework{} obtains good results in terms of performance and accuracy.

We first present the verification results of masked hardware cipher accelerators, which highlights the concordance of the verification verdicts with the state of the art, and which are obtained in a shorter time. Second, we present the formal leakage analysis of masked software running on five different CPUs, which demonstrates the scalability and versatility of the approach, while verifying in particular proven leakage-free versions for all benchmarks. Finally, for two CPUs, we experimentally validate the results obtained with \framework{} on two programs, by performing real power measurements of the same programs on the same targets. We show that when \framework{} concludes that no secret leakage exists in the \lmd{RR 1-sw-probing model}, no such leakage can be observed in practice.

The experimental setup for all the experiments is an AMD Ryzen 9 7945HX computer with 92GB of RAM and 512GB of SSD swap memory, from the Dalek cluster~\cite{cassagne2025dalekunconventionalenergyawareheterogeneous}.

\subsection{Masked Hardware Accelerators}

Our analysis of already verified ciphers is performed on 3 main ciphers, from which we decline 8 benchmarks. We summarise the results obtained using \framework{} on these benchmarks in Table~\ref{tb:res-implem-hw}. We also reproduce the verification on the same experimental setup using \coco{} which is, to the best of our knowledge, the only tool able to formally verify stateful ciphers in a leakage model considering transitions, glitches and stability. The hardware ciphers presented were all designed to be secure in bit probing models, thus, the verification performed is in the \lmd{RR 1-bit-probing} model.

\begin{table}
    \small
    \centering
    \resizebox{\linewidth}{!}{\begin{tabular}{l|r|l@{}rc||l@{}rc|c|c}
        \hline
        \multirow{2}{*}{Masked hardware} & \multirow{2}{*}{\makecell{Verif/Total\\ Cycles}} & \multicolumn{3}{c||}{aLEAKator} & \multicolumn{3}{c|}{\coco{}} & \prover{} & \prolead{} \\
        & & \multicolumn{2}{c}{Verdict} & RAM (GB) & \multicolumn{2}{c}{Verdict} & RAM (GB) & Verdict & Verdict \\ \hline
        \tool{SBox Present-TI NU} & 2/2    & \xmark  & <1s  & 0.01 & \xmark  & <1s   & 0.03 & \xmark & \xmark \\
        \tool{SBox Present-TI U}  & 2/2    & \cmark  & <1s  & 0.01 & \xmark* &  <1s  & 0.03 & \cmark & \cmark \\ \hline
        \tool{Present-TI NU}      & 3/544  & \xmark  & <1s  & 0.01 & -       &       & -    & - & \xmark \\
        \tool{Present-TI U}       & 3/544  & \cmark  & <1s  & 0.01 & -       &       & -    & - & \cmark \\
        \tool{AES-DOM}            & 21/216 & \cmark  & 30s  & 12.3 & \cmark  & 26m30 & 3    & - & - \\
        \tool{Prince-TI}          & 3/25   & \xmark  & 4.5s & 2.2  & \xmark  & 4.6s  & 0.4  & - & - \\
        \tool{Prince-TI-Coco}     & 3/25   & \cmark  & 5s   & 2.7  & \cmark  & 29s   & 0.4  & - & - \\
        \tool{Prince-TI-Coco}     & 5/25   & \xmark* & 16s  & 8.0  & \cmark  & 3m45  & 1.7  & - & - \\ \hline
    \end{tabular}}
    \caption{Verification results obtained by \framework{}, \coco{}, \prover{} and \prolead{}.}
    \label{tb:res-implem-hw}
\end{table}

\noindent\textbf{\tool{Present}} is a lightweight block cipher primarily designed for hardware implementations, consisting of 31 rounds. Shahmirzadi \textit{et al.}~\cite{TCHES:ShaMor21a} proposed \tool{Present-TI}, a TI-masked version of the \tool{Present} cipher. Two variants of the cipher are available: one using a non-uniform (\tool{NU}) SBox~\cite{AC:EGMP17}, and the other using a uniform (\tool{U}) SBox~\cite{JC:PMKLWL11}. As shown by \prover{}~\cite{prover}, \coco{} experiment false-positives on both SBox versions while \prover{} can successfully verify that the uniform version is the only secure one. \prolead{} verified both versions as well as the full ciphers using them. \framework{} corrects the eventual false-positives by using enumeration, allowing to confirm that the uniform SBox is secure and the non-uniform one is not. We also verify the three first cycles of the full cipher to obtain concordant results with \prolead{} but from cycle 3 onward, \framework{} relies on enumeration. The fourth cycle is verified within 26 minutes with the same amount of memory usage. The results of \coco{} for the full cipher are not given here as it is already providing false-positive results for the SBox.

As \prover{} does not handle circuits in which inputs can vary over time, results are not given for \tool{Present}, \tool{AES-DOM} and \tool{Prince}. Also, the version of \tool{AES-DOM} and \tool{Prince} verified by \prolead{} are not the same implementations and thus are not given either.

\noindent\textbf{\tool{AES-DOM}} is a DOM-masked VHDL open source implementation of the AES-128 cipher proposed by Gross \textit{et al}.~\cite{grossDomainOrientedMaskingCompact2016} with a configurable security order. It was previously verified, in its non perfectly interleaved version but with an eight stage SBox version, by \prolead{} to be \lmd{(1,1) 1-probing secure} (it is no longer available in the benchmarks of \tool{Prolead-V3} as of the date of submission). The perfectly interleaved version with seven stage SBoxes was verified by \coco{} up to the 21$^{\mathrm{st}}$ cycle. We use \framework{} to reproduce the results with the same configuration as \coco{} in Table~\ref{tb:res-implem-hw}. The accelerator is fed with 38 random bits per cycle. The shares of the key and plaintext are concatenated byte by byte and are provided to the circuit in 16 cycles. When the verification is pushed further, \coco{} times out while computing the 22$^{\mathrm{nd}}$ cycle while \framework{} can verify up to the 23$^{\mathrm{th}}$ cycle. We note that on this benchmark, it is more than 50 times faster than \coco{}, with only 5.5 times more RAM usage.

\noindent\textbf{\tool{Prince}} is a lightweight block cipher, smaller than \tool{Present}, requiring only 11 rounds to compute.  Božilov \textit{et al.}~\cite{JCEng:BozKneNik22} proposed \tool{Prince-TI}, a TI-masked implementation of the \tool{Prince} cipher. It was proven unsecure by \coco{} due to a glitchy control signal, that \framework{} identifies as well on the same multiplexer. A fixed version proposed by \coco{}'s authors was verified up to cycle 7. \framework{} is not able to conclude at cycle 5 and as enumeration would not be tractable, we interpret this result as a false-positive following \coco{}'s result. Still, \framework{} reproduces the secure verification verdict for the three first cycles in only the sixth of the time needed for \coco{}. 

On summary, \framework{} is able to reproduce known verification results, while still suffering from eventual false-positive results. On non trivial circuits, \framework{} is up to 50 times faster than \coco{}, for a higher memory footprint.

\subsection{Masked Software on CPUs}

This section details the results obtained while verifying software ciphers on five CPUs. We first describe the target CPUs and benchmarks. 
We then present the complete verification results, and we end this section by showing, using power traces measured on real hardware, that using \framework{} allowed to remove all observed real-world leakages.

\subsubsection{Hardware Targets Configuration}

To showcase the versatility of \framework{}, we apply our method on five CPUs with memory subsystems, that we present in the following.

\noindent\textbf{\ibex{}} \footnote{Available at \url{https://github.com/lowRISC/ibex}.} is a CPU core implementing the open source 32-bit RISC-V ISA. Initially developed as part of the PULP platform, it is now maintained and developed by lowRISC. This core is highly configurable and implements a two or three stage pipeline (with or without write-back stage). It can also implement various multipliers and optionally a branch target ALU. The programs verified on the \ibex{} are verified in the three stage pipeline version. Experiments are made with the ibex simple system, an example bus and memory subsystem from the \ibex{} project.

\noindent\textbf{\cocoibex{}}~\cite{USENIX:GHPMB21}\footnote{Available at \url{https://github.com/isec-tugraz/coco-ibex}.} is a fork of the \ibex{} core secured by the authors of \coco{}. It features a read/write gated register file, a secure RAM, and gating mechanisms for the multiplier, shifter, adder, as well as control and status registers units. Each protection can be independently activated but we verify programs with all the protections enabled. Verification on this core uses the same bus as the \ibex{} core for accessing the unprotected ROM. We place our verification in the configuration of the authors of \coco{}~\cite{USENIX:GHPMB21}, with the two-stage pipeline configuration without a write-back stage. \cocoibex{} implements a secure memory using fully mapped flip-flop cells. As a result, verifications for this CPU go deeper in the implementation of memories than other CPUs that use off-the-shelf SRAM cuts. We do not run programs such as the \tool{AES-Herbst} which requires more than 512 bytes of RAM, since increasing RAM to 4096 bytes would double the \cocoibex{} core’s gate count. The version of \cocoibex{} used in our experiments includes fixes to hardware bugs that prevented the use of certain common instructions, such as \tool{c.jal}, which are required by our benchmarks. As \cocoibex{} implements countermeasures against side-channel attacks, some common leakage sources are mitigated by the hardware for this core. We note that following the authors’ recommendations regarding the memory placement of shares is necessary to achieve non-leaking memory operations. Nevertheless, programs still need to be secured and verified, as the micro-architecture does not eliminate all secret leakages.

\noindent\textbf{\cv{}}~\cite{cv32e40p}\footnote{Available at \url{https://github.com/openhwgroup/cv32e40p}.} is a small and efficient open source 32-bit RISC-V processor with a 4-stage pipeline aimed at performance and energy efficiency. Originally maintained by the PULP platform; it is now maintained and developed by the OpenHW group. Experiments were made using the same memory subsystem as the \ibex{}.

\noindent\textbf{\cmthree{}} and \textbf{\cmfour{}} are two general purpose 32-bit three stage pipeline ARMv7-M CPU designed for high-performance and low-cost platforms. Their HDL implementations are made available to us by Arm through the Arm Academic Access (AAA) program. An example AHB-Lite bus is provided with the cores, which we used for our verifications. We disabled tracing, debug and JTAG options for the experiments.

\subsubsection{Verified Programs}

\label{sc:results_sw}

For each program, two versions are provided, a non-secure and a secure one. Each secure version has been obtained by iteratively removing secret leakages using \framework{}, until reaching a non-leaking result in the desired leakage model.

\noindent\textbf{\tool{DOM-AND}}~\cite{grossDomainOrientedMaskingCompact2016} is an implementation of the \textsc{and} function, masked with the \tool{DOM} masking scheme. \tool{DOM-AND-nsec}, the non-secure implementation, is an assembly version which has been proven secure when considering transitions at the GPR level. The \tool{DOM-AND-sec-rr} implementations were designed to be secure in the \lmd{RR 1-sw-probing model}.

\noindent\textbf{\tool{Secmult}}~\cite{CHES:RivPro10} is a masked implementation of the multiplication in GF($2^{8}$). For the \tool{Cortex-M3/M4} cores, the non-secure implementation, \tool{Secmult-nsec}, is a \texttt{-O3} compiled version of the algorithm. For RISC-V processors, it refers to a manually-secured version in the value without glitches leakage model. The secure versions, named \tool{Secmult-sec-rr}, refer to versions made to be secure in the \lmd{RR 1-sw-probing model}.

\noindent\textbf{\tool{AES-Herbst}}~\cite{ACNS:HerOswMan06} is a masked tabulated implementation of the AES cipher using 6 masks, including the key schedule. The non-secure version, \tool{AES-Herbst-nsec}, is a \texttt{-O3} compiled version of the cipher. As providing a secure version of the cipher in assembly would be a long task, the secure version, \tool{AES-Herbst-sec-v}, is a \tool{C}-level hardened implementation made to be secure in the \lmd{(0,0) 1-sw-probing model}. We still provide results of how \framework{} performs when verifying this cipher in the \lmd{RR 1-sw-probing model} but we note that leaking implementations are usually longer and more costly to verify than non leaking ones, while it is usually desirable to stop at the first leakage found.

This implementation is tabulated using symbolic indexes. To tackle this issue, we used the overloading feature of the memory system for each CPU, so that the symbolic expression given as array index is used to determine the result of the memory access. In particular, the masking scheme of this cipher requires to compute a table \tool{SBox'} which is initialised as \tool{SBox'[i $\oplus$ m] = SBox[i] $\oplus$ m'} for all \tool{i} values $\in$ [0; 255], with \tool{m} and \tool{m'} being two masks in the scheme. Upon this access, this expression transformation can only be done with this overloading feature, and this initialisation information.

\subsubsection{Software Verification Results}

The verification results are presented in Table~\ref{tb:results-sw}. For each CPU, we present each version of the three programs, except for \tool{AES-Herbst} on the \cocoibex{}, as explained above. 

\begin{table}
\caption{Detailed \framework{} verification results.}
\label{tb:results-sw}
\small
\begin{tabular}{|l|@{}c@{}|@{}c@{}|@{}c@{}|@{}c@{}|r|r|r|r|r|c|}
\hline
 & SW & Ver. & F & \multicolumn{1}{c|}{\begin{tabular}[c]{@{}c@{}}Leak.\\ Model\end{tabular}} & \multicolumn{1}{c|}{Cycles} & \multicolumn{1}{c|}{\begin{tabular}[c]{@{}c@{}}Expr. to\\ Verify\end{tabular}} & \multicolumn{1}{c|}{\begin{tabular}[c]{@{}c@{}}Verified\\ Expr.\end{tabular}} & \multicolumn{1}{c|}{\begin{tabular}[c]{@{}c@{}}Leak.\\ Cycles\end{tabular}} & \multicolumn{1}{@{}c@{}|}{Time} & \multicolumn{1}{c|}{\begin{tabular}[c]{@{}c@{}}RAM\\ (GB)\end{tabular}} \\ \hline \hline
\multirow{9}{*}{\rotatebox[origin=c]{90}{Cortex-M3}}
  & \multirow{3}{*}{\hphantom{\small .}\tool{DOM-AND}\hphantom{\small .}}
    & \multirow{2}{*}{\xsec} & \multirow{2}{*}{\ref{fig:dom-and-m3-unsecure}}
      & Value & 48 & 630 & 630 & \textcolor{cnoleak}{0} & 1.8s & 0.6 \\ \cdashline{5-11}
      &&&& RR & 48 & 1 092 & 733 & \textcolor{cleak}{33} & 3.8s & 1.0 \\ \cdashline{3-11}
    && \hphantom{\small .}\tool{sec-rr}\hphantom{\small .} & \ref{fig:dom-and-m3-secure} & RR & 67 & 1 095 & 779 & \textcolor{cnoleak}{0} & 4.5s & 1.0 \\ \cline{2-11}

  & \multirow{3}{*}{\tool{Secmult}}
    & \multirow{2}{*}{\xsec} & \multirow{2}{*}{\hphantom{\tiny .}\ref{fig:secmult-m3-unsecure}\hphantom{\tiny .}}
      & Value & 209 & 2 916 & 2 916 & \textcolor{cleak}{53} & 5.1s & 0.5 \\ \cdashline{5-11}
      &&&& RR & 209 & 4 591 & 3 469 & \textcolor{cleak}{185} & 13.5s & 1.4 \\ \cdashline{3-11}
    && \vsecrr & \ref{fig:secmult-m3-secure} & RR & 660 & 4 494 & 3 186 & \textcolor{cnoleak}{0} & 34.0s & 1.7 \\ \cline{2-11}

  & \multirow{3}{*}{\makecell{\tool{AES-}\\\tool{Herbst}}}
      & \xsec && Value & 6 924 & 74 707 & 74 707 & \textcolor{cleak}{3 050} & 3m46 & 37.1 \\ \cdashline{3-11}
      && \multirow{2}{*}{\vsec}
        && Value & 10 113 & 90 053 & 90 053 & \textcolor{cnoleak}{0} & 5m04 & 43.7 \\ \cdashline{4-11}
        &&&& RR & 10 113 & 192 812 & 148 409 & \textcolor{cleak}{8 329} & 18m53 & 208.0 \\ \hline\hline
\multirow{9}{*}{\rotatebox[origin=c]{90}{Cortex-M4}}
  & \multirow{3}{*}{\tool{DOM-AND}}
    & \multirow{2}{*}{\xsec} & \multirow{2}{*}{\ref{fig:dom-and-m4-unsecure}}
      & Value & 48 & 649 & 649 & \textcolor{cnoleak}{0} & 1.9s & 0.6 \\ \cdashline{5-11}
      &&&& RR & 48 & 1 112 & 762 & \textcolor{cleak}{33} & 4.3s & 1.1 \\ \cdashline{3-11}
    && \vsecrr & \ref{fig:dom-and-m4-secure} & RR & 67 & 1 112 & 799 & \textcolor{cnoleak}{0} & 5.0s & 1.0 \\ \cline{2-11}

  & \multirow{3}{*}{\tool{Secmult}}
    & \multirow{2}{*}{\xsec} & \multirow{2}{*}{\ref{fig:secmult-m4-unsecure}}
      & Value & 203 & 2 998 & 2 998 & \textcolor{cleak}{36} & 5.7s & 0.5 \\ \cdashline{5-11}
      &&&& RR & 203 & 4 669 & 3 419 & \textcolor{cleak}{179} & 14s & 1.4 \\ \cdashline{3-11}
    && \vsecrr & \ref{fig:secmult-m4-secure} & RR & 660 & 4 680 & 3 255 & \textcolor{cnoleak}{0} & 39s & 1.8 \\ \cline{2-11}

    & \multirow{3}{*}{\makecell{\tool{AES-}\\\tool{Herbst}}}
      & \xsec && Value & 6 924 & 77 227 & 77 227 & \textcolor{cleak}{3 050} & 4m13 & 38.0 \\ \cdashline{3-11}
      & & \multirow{2}{*}{\vsec}
        && Value & 10 113 & 93 279 & 93 279 & \textcolor{cnoleak}{0} & 5m48 & 45.3 \\ \cdashline{4-11}
        &&&& RR & 10 113 & 194 470 & 147 766 & \textcolor{cleak}{8 329} & 20m14 & 211.5 \\ \hline\hline

\multirow{9}{*}{\rotatebox[origin=c]{90}{Ibex}}
  & \multirow{3}{*}{\tool{DOM-AND}}
    & \multirow{2}{*}{\xsec} &
      & Value & 37 & 705 & 705 & \textcolor{cnoleak}{0} & 5.1s & 3.4 \\ \cdashline{4-11}
      &&&& RR & 37 & 881 & 350 & \textcolor{cleak}{18} & 2.7s & 1.0 \\ \cdashline{3-11}
    && \vsecrr && RR & 58 & 833 & 355 & \textcolor{cnoleak}{0} & 2.1s & 0.7 \\ \cline{2-11}

  & \multirow{3}{*}{\tool{Secmult}}
    & \multirow{2}{*}{\xsec} &
      & Value & 638 & 12 703 & 12 703 & \textcolor{cnoleak}{0} & 11.4s & 2.8 \\ \cdashline{4-11}
      &&&& RR & 638 & 17 609 & 4 527 & \textcolor{cleak}{236} & 21.3s & 4.5 \\ \cdashline{3-11}
    && \vsecrr && RR & 646 & 17 270 & 4 427 & \textcolor{cnoleak}{0} & 20.5s & 4.3 \\ \cline{2-11}

  & \multirow{3}{*}{\makecell{\tool{AES-}\\\tool{Herbst}}}
      & \xsec && Value & 6 294 & 159 261 & 159 261 & \textcolor{cleak}{2 963} & 14m37 & 203.1 \\ \cdashline{3-11}
      && \multirow{2}{*}{\vsec} &
        & Value & 8 869 & 185 508 & 185 508 & \textcolor{cnoleak}{0} & 21m41 & 226.8 \\ \cdashline{4-11}
        &&&& RR & 8 869 & 230 463 & 90 050 & \textcolor{cleak}{6 651} & 22m08 & 301.0 \\ \hline \hline

\multirow{9}{*}{\rotatebox[origin=c]{90}{CV32E40P}}
  & \multirow{3}{*}{\tool{DOM-AND}}
    & \multirow{2}{*}{\xsec} &
      & Value & 37 & 785 & 785 & \textcolor{cnoleak}{0} & 10.5s & 9.9 \\ \cdashline{4-11}
      &&&& RR & 37 & 860 & 299 & \textcolor{cleak}{17} & 5.8s & 1.1 \\ \cdashline{3-11}
    && \vsecrr && RR & 60 & 822 & 231 & \textcolor{cnoleak}{0} & 7.5s & 0.8 \\ \cline{2-11}

  & \multirow{3}{*}{\tool{Secmult}}
    & \multirow{2}{*}{\xsec} &
      & Value & 639 & 13 824 & 13 824 & \textcolor{cnoleak}{0} & 15.8s & 3.4 \\ \cdashline{4-11}
      &&&& RR & 639 & 12 461 & 2 886 & \textcolor{cleak}{234} & 1m14 & 5.0 \\ \cdashline{3-11}
    && \vsecrr && RR & 651 & 12 228 & 2 783 & \textcolor{cnoleak}{0} & 1m15 & 4.8 \\ \cline{2-11}

  & \multirow{3}{*}{\makecell{\tool{AES-}\\\tool{Herbst}}}
      & \xsec && Value & 6 294 & 176 150 & 176 150 & \textcolor{cleak}{2 762} & 21m38 & 267.6 \\ \cdashline{3-11}
      && \multirow{2}{*}{\vsec} &
        & Value & 8 869 & 199 405 & 199 405 & \textcolor{cnoleak}{0} & 43m44 & 423.5 \\ \cdashline{4-11}
        &&&& RR & 8 869 & 207 301 & 86 259 & \textcolor{cleak}{6 705} & 39m01 & 333.7 \\ \hline \hline

\multirow{6}{*}{\rotatebox[origin=c]{90}{\cocoibex{}}}
  & \multirow{3}{*}{\tool{DOM-AND}}
    & \multirow{2}{*}{\xsec} &
      & Value & 51 & 537 & 537 & \textcolor{cnoleak}{0} & 1.5s & 0.5 \\ \cdashline{4-11}
      &&&& RR & 51 & 1 223 & 651 &\textcolor{cleak}{9} & 6.8s & 3.5 \\ \cdashline{3-11}
    && \vsecrr && RR & 61 & 1 194 & 961 & \textcolor{cnoleak}{0} & 7.6s & 3.8 \\ \cline{2-11}

  & \multirow{3}{*}{\tool{Secmult}}
    & \multirow{2}{*}{\xsec} &
      & Value & 724 & 2 033 & 2 033 & \textcolor{cnoleak}{0} & 34.3s & 8.3 \\ \cdashline{4-11}
      &&&& RR & 724 & 4 639 & 3 583 & \textcolor{cleak}{10} & 1m30 & 43.6 \\ \cdashline{3-11}
    && \vsecrr && RR & 739 & 4 622 & 3 533 & \textcolor{cnoleak}{0} & 1m30 & 43.7 \\ \hline \hline

 & SW & Ver. & F & \multicolumn{1}{c|}{\begin{tabular}[c]{@{}c@{}}Leak.\\ Model\end{tabular}} & \multicolumn{1}{c|}{Cycles} & \multicolumn{1}{c|}{\begin{tabular}[c]{@{}c@{}}Expr. to\\ Verify\end{tabular}} & \multicolumn{1}{c|}{\begin{tabular}[c]{@{}c@{}}Verified\\ Expr.\end{tabular}} & \multicolumn{1}{c|}{\begin{tabular}[c]{@{}c@{}}Leak.\\ Cycles\end{tabular}} & \multicolumn{1}{@{}c@{}|}{Time} & \multicolumn{1}{c|}{\begin{tabular}[c]{@{}c@{}}RAM\\ (GB)\end{tabular}} \\ \hline
\end{tabular}
\end{table}

As the verification manager excludes trivial verifications and cached results, the number of expressions sent to the verifier is far smaller than the number of wires of the circuit multiplied by the number of cycles. To better illustrate the effect of our over-approximation detailed in Section~\ref{sc:wires-to-verify}, we give two columns for the number of expression sets sent to the verifier. The first is \emph{Expr. to Verify}, the number of expression sets that should be sent to the verifier when not using the over-approximation (while still using the cache and removing trivial expression sets). The second, \emph{Verified Expr.}, is the number of expression sets sent to the verifier in practice, using the over-approximation. We emphasise that the difference between these two columns is the number of unique verifications which were not performed thanks to the over-approximation. This difference is only apparent for verifications in the \lmd{RR 1-sw-probing model} as all wires must be verified when considering the \lmd{(0,0) 1-sw-probing model}.

The \emph{Leak. Cycles} column reports the number of cycles, among the cycles of the execution (column \emph{Cycles}), for which at least one expression set cannot be proven non-leaking by the verifier.

We also report the time and RAM consumption for each program verification.
We note that different CPUs can cause different RAM usage in simulation. This behaviour can be explained by the gating of some operators, like the multiplier or the shifter in 
the Ibex and CV32E40P for example. Non-gated operators tend to generate more symbolic expressions and, depending on the operator, expressions that are harder to verify. The RAM usage difference is also due to the number of gates, the number of simulated cycles and the number of already found leakages. On complex programs such as \tool{AES-Herbst} RAM consumption is high but, as our experiments have been performed using swap, we consider this usage acceptable.

The various optimisations are an essential part of \framework{}. When running \tool{AES-Herbst-sec-v} on the Cortex-M3, over 80 millions trivial verifications are excluded, then around 2 million cache hits occur, allowing for a significant reduction of the verification time and RAM consumption. The over-approximation allows for the reduction by an average factor of 2.2 and up to a factor of 4.4 -- for the CV32E40P's \tool{Secmult} -- of the number of unique expression sets to verify.

\framework{} is efficient as it can verify ciphers requiring thousands of computation cycles, such as \tool{AES-Herbst}, within the \lmd{RR 1-probing model} in 19 to 39 minutes, depending on the targeted CPU. When operating under a value leakage model, the verification can be completed in as little as five minutes on a Cortex-M3 processor.

\begin{mdframed}
The major results from the verification are that, for the \tool{sec-rr} and \tool{sec-v} versions of the program for which there is no leaking cycle, the implementations are proven secure on the corresponding targets in their respective leakage models. To the best of our knowledge, this is the first verification in both the \lmd{(0,0) 1-(sw/bit)-probing model} and the \lmd{RR 1-(sw/bit)-probing model} of a secure \tool{AES-Herbst} in the \lmd{(0,0) 1-probing model} for the four processors. This paves the way for, if an assembly implementation was to be made, a secure implementation for those CPUs in the \lmd{RR 1-sw-probing model}.
\end{mdframed}

\subsubsection{Comparison with Real Power Measurements}

We validate the results produced by \framework{} by performing power measurements on real CPUs, on the programs \tool{DOM-AND} and \tool{Secmult}. For each program, both the \tool{nsec} and the \tool{sec-rr} versions have been considered. Measurements were made using the \tool{ChipWhisperer} CW1200 with the \tool{STM32F1} and \tool{STM32F3} target boards, embedding respectively a Cortex-M3 and a Cortex-M4 core~\cite{COSADE:OFlChe14}.

\noindent\textbf{Comparison of \framework{} verdict with $t$-test measurements.} The leakages reported by \framework{} happen in many central parts of the CPU including the Load-Store Unit, the register bank, ALU ports and internal ALU signals. Figures~\ref{fig:dom-and-m3},~\ref{fig:dom-and-m4},~\ref{fig:secmult-m3} and \ref{fig:secmult-m4} represent the evolution of the $t$-test value for each secret input for 500,000 traces, and for the entire duration of the program, with 4 samples per cycle. Each case is reported in Table~\ref{tb:results-sw} in the "F" column. The dotted red line represents the 4.5 value for the $t$-test, above which a secret is considered to be leaking. These figures show that all non secure versions exhibit very high leakage values, while all secure versions do not exhibit any. These results confirm that when \framework{} reports no secret leakage, no such leakages are observed in practice.

\newlength{\figheigth}
\setlength{\figheigth}{4.5cm}

\begin{figure}
    \centering
    \begin{subfigure}{0.51\textwidth}
    \caption{\tool{DOM-AND-nsec}}
    \label{fig:dom-and-m3-unsecure}
    \begin{tikzpicture}[remember picture]
    \pgfplotstablegetrowsof{exps/data_m3/dom_and_gprt/m3_dom_and_gprt_e_0.csv}\pgfmathsetmacro\nrows{\pgfmathresult+15}
    \begin{axis}[
        width=6.98cm,
        height=\figheigth,
        table/col sep=comma,
        xlabel={time ($sample$)},
        ylabel={$|t$-test$|$},
        xmin=10, 
        xmax=\nrows,
        x label style={inner ysep=0pt, outer ysep=0pt},
        title style={inner ysep=0pt, outer ysep=0pt},
        enlargelimits=false,
        axis x line=bottom,
        axis y line=left,
        legend style={nodes={scale=0.75, transform shape}},
    ]
        \addplot [Paired-7] table [y expr={abs(\thisrowno{1})}] {exps/data_m3/dom_and_gprt/m3_dom_and_gprt_e_0.csv};
        \addlegendentry{Input \texttt{a}}
        \addplot [Paired-1] table [y expr={abs(\thisrowno{1})}] {exps/data_m3/dom_and_gprt/m3_dom_and_gprt_e_1.csv};
        \addlegendentry{Input \texttt{b}}
        \draw[Paired-5, dashed, line width=1.2pt] (axis cs:0,4.5) -- (axis cs:\nrows,4.5);
    \end{axis}
    \end{tikzpicture}
    \end{subfigure}\hfill{}
    \begin{subfigure}{0.48\textwidth}
    \caption{\tool{DOM-AND-sec-rr}}
    \label{fig:dom-and-m3-secure}
    \begin{tikzpicture}[remember picture]
    \pgfplotstablegetrowsof{exps/data_m3/dom_and_RR/dom_and_twg_e_0.csv}\pgfmathsetmacro\nrows{\pgfmathresult+10}
    \begin{axis}[
        width=6.98cm,
        height=\figheigth,
        table/col sep=comma,
        xlabel={time ($sample$)},
        ylabel={$|t$-test$|$},
        xmin=10, 
        xmax=\nrows,
        ymax=6.5,
        x label style={inner ysep=0pt, outer ysep=0pt},
        title style={inner ysep=0pt, outer ysep=0pt},
        enlargelimits=false,
        axis x line=bottom,
        axis y line=left,
        legend style={nodes={scale=0.75, transform shape}},
    ]
        \addplot [Paired-7] table [y expr={abs(\thisrowno{1})}] {exps/data_m3/dom_and_RR/dom_and_twg_e_0.csv};
        \addlegendentry{Input \texttt{a}}
        \addplot [Paired-1] table [y expr={abs(\thisrowno{1})}] {exps/data_m3/dom_and_RR/dom_and_twg_e_1.csv};
        \addlegendentry{Input \texttt{b}}
        \draw[Paired-5, dashed, line width=1.2pt] (axis cs:0,4.5) -- (axis cs:\nrows,4.5);
    \end{axis}
    \end{tikzpicture}
    \end{subfigure}
    \caption{$t$-test values for the unsecure (\subref{fig:dom-and-m3-unsecure}) and secure  (\subref{fig:dom-and-m3-secure}) \tool{DOM-AND} on the Cortex-M3.}
    \label{fig:dom-and-m3}
\end{figure}

\begin{figure}
    \centering
    \begin{subfigure}{0.51\textwidth}
    \caption{\tool{DOM-AND-nsec}}
    \label{fig:dom-and-m4-unsecure}
    \begin{tikzpicture}[remember picture]
    \pgfplotstablegetrowsof{exps/data_m4/dom_and_gprt/dom_and_gprt_e_0.csv}\pgfmathsetmacro\nrows{\pgfmathresult+10}
    \begin{axis}[
        width=6.98cm,
        height=\figheigth,
        table/col sep=comma,
        xlabel={time ($sample$)},
        ylabel={$|t$-test$|$},
        xmin=10, 
        xmax=\nrows,
        x label style={inner ysep=0pt, outer ysep=0pt},
        title style={inner ysep=0pt, outer ysep=0pt},
        enlargelimits=false,
        axis x line=bottom,
        axis y line=left,
        x tick label style={/pgf/number format/.cd, set thousands separator={}},
        legend style={nodes={scale=0.75, transform shape}},
    ]
        \addplot [Paired-7] table [y expr={abs(\thisrowno{1})}] {exps/data_m4/dom_and_gprt/dom_and_gprt_e_0.csv};
        \addlegendentry{Input \texttt{a}}
        \addplot [Paired-1] table [y expr={abs(\thisrowno{1})}] {exps/data_m4/dom_and_gprt/dom_and_gprt_e_1.csv};
        \addlegendentry{Input \texttt{b}}
        \draw[Paired-5, dashed, line width=1.2pt] (axis cs:0,4.5) -- (axis cs:\nrows,4.5);
    \end{axis}
    \end{tikzpicture}
    \end{subfigure}\hfill{}
    \begin{subfigure}{0.48\textwidth}
    \caption{\tool{DOM-AND-sec-rr}}
    \label{fig:dom-and-m4-secure}
    \begin{tikzpicture}[remember picture]
    \pgfplotstablegetrowsof{exps/data_m4/dom_and_RR/dom_and_twg_ncwb_e_0.csv}\pgfmathsetmacro\nrows{\pgfmathresult+10}
    \begin{axis}[
        width=6.98cm,
        height=\figheigth,
        table/col sep=comma,
        xlabel={time ($sample$)},
        ylabel={$|t$-test$|$},
        xmin=10, 
        xmax=\nrows,
        ymax=6.5,
        x label style={inner ysep=0pt, outer ysep=0pt},
        title style={inner ysep=0pt, outer ysep=0pt},
        enlargelimits=false,
        axis x line=bottom,
        axis y line=left,
        x tick label style={/pgf/number format/.cd, set thousands separator={}},
        legend style={nodes={scale=0.75, transform shape}},
    ]
        \addplot [Paired-7] table [y expr={abs(\thisrowno{1})}] {exps/data_m4/dom_and_RR/dom_and_twg_ncwb_e_0.csv};
        \addlegendentry{Input \texttt{a}}
        \addplot [Paired-1] table [y expr={abs(\thisrowno{1})}] {exps/data_m4/dom_and_RR/dom_and_twg_ncwb_e_1.csv};
        \addlegendentry{Input \texttt{b}}
        \draw[Paired-5, dashed, line width=1.2pt] (axis cs:0,4.5) -- (axis cs:\nrows,4.5);
    \end{axis}
    \end{tikzpicture}
    \end{subfigure}
    \caption{$t$-test values for the unsecure (\subref{fig:dom-and-m4-unsecure}) and secure(\subref{fig:dom-and-m4-secure}) \tool{DOM-AND} on the Cortex-M4.}
    \label{fig:dom-and-m4}
\end{figure}

\begin{figure}
    \centering
    \begin{subfigure}{0.51\textwidth}
    \caption{\tool{Secmult-nsec}}
    \label{fig:secmult-m3-unsecure}
    \begin{tikzpicture}[remember picture]
    \pgfplotstablegetrowsof{exps/data_m3/secmult_unsecure/m3_secmult_e_0.csv}\pgfmathsetmacro\nrows{\pgfmathresult+15}
    \begin{axis}[
        width=6.98cm,
        height=\figheigth,
        table/col sep=comma,
        xlabel={time ($sample$)},
        ylabel={$|t$-test$|$},
        xmin=10, 
        xmax=\nrows,
        x label style={inner ysep=0pt, outer ysep=0pt},
        title style={inner ysep=0pt, outer ysep=0pt},
        enlargelimits=false,
        axis x line=bottom,
        axis y line=left,
        x tick label style={/pgf/number format/.cd, set thousands separator={}},
        legend style={nodes={scale=0.75, transform shape}},
    ]
        \addplot [Paired-7] table [y expr={abs(\thisrowno{1})}] {exps/data_m3/secmult_unsecure/m3_secmult_e_0.csv};
        \addlegendentry{Input \texttt{a}}
        \addplot [Paired-1] table [y expr={abs(\thisrowno{1})}] {exps/data_m3/secmult_unsecure/m3_secmult_e_1.csv};
        \addlegendentry{Input \texttt{b}}
        \draw[Paired-5, dashed, line width=1.2pt] (axis cs:0,4.5) -- (axis cs:\nrows,4.5);
    \end{axis}
    \end{tikzpicture}
    \end{subfigure}\hfill{}
    \begin{subfigure}{0.48\textwidth}
    \caption{\tool{Secmult-sec-rr}}
    \label{fig:secmult-m3-secure}
    \begin{tikzpicture}[remember picture]
    \pgfplotstablegetrowsof{exps/data_m3/secmult_RR/m3_secmult_twg_e_0.csv}\pgfmathsetmacro\nrows{\pgfmathresult+125}
    \begin{axis}[
        width=6.98cm,
        height=\figheigth,
        table/col sep=comma,
        xlabel={time ($sample$)},
        ylabel={$|t$-test$|$},
        xmin=10, 
        xmax=\nrows,
        ymax=6.5,
        x label style={inner ysep=0pt, outer ysep=0pt},
        title style={inner ysep=0pt, outer ysep=0pt},
        enlargelimits=false,
        axis x line=bottom,
        axis y line=left,
        x tick label style={/pgf/number format/.cd, set thousands separator={}},
        legend style={nodes={scale=0.75, transform shape}},
    ]
        \addplot [Paired-7] table [y expr={abs(\thisrowno{1})}] {exps/data_m3/secmult_RR/m3_secmult_twg_e_0.csv};
        \addlegendentry{Input \texttt{a}}
        \addplot [Paired-1] table [y expr={abs(\thisrowno{1})}] {exps/data_m3/secmult_RR/m3_secmult_twg_e_1.csv};
        \addlegendentry{Input \texttt{b}}
        \draw[Paired-5, dashed, line width=1.2pt] (axis cs:0,4.5) -- (axis cs:\nrows,4.5);
    \end{axis}
    \end{tikzpicture}
    \end{subfigure}
    \caption{$t$-test values for the unsecure (\subref{fig:secmult-m3-unsecure}) and secure (\subref{fig:secmult-m3-secure}) \tool{Secmult} on the Cortex-M3.}
    \label{fig:secmult-m3}
\end{figure}

\begin{figure}
    \centering
    \begin{subfigure}{0.51\textwidth}
    \caption{\tool{Secmult-nsec}}
    \label{fig:secmult-m4-unsecure}
    \begin{tikzpicture}[remember picture]
    \pgfplotstablegetrowsof{exps/data_m4/secmult_unsecure/secmult_e_0.csv}\pgfmathsetmacro\nrows{\pgfmathresult+15}
    \begin{axis}[
        width=6.98cm,
        height=\figheigth,
        table/col sep=comma,
        xlabel={time ($sample$)},
        ylabel={$|t$-test$|$},
        xmin=10, 
        xmax=\nrows,
        x label style={inner ysep=0pt, outer ysep=0pt},
        title style={inner ysep=0pt, outer ysep=0pt},
        enlargelimits=false,
        axis x line=bottom,
        axis y line=left,
        x tick label style={/pgf/number format/.cd, set thousands separator={}},
        legend style={nodes={scale=0.75, transform shape}},
    ]
        \addplot [Paired-7] table [y expr={abs(\thisrowno{1})}] {exps/data_m4/secmult_unsecure/secmult_e_0.csv};
        \addlegendentry{Input \texttt{a}}
        \addplot [Paired-1] table [y expr={abs(\thisrowno{1})}] {exps/data_m4/secmult_unsecure/secmult_e_1.csv};
        \addlegendentry{Input \texttt{b}}
        \draw[Paired-5, dashed, line width=1.2pt] (axis cs:0,4.5) -- (axis cs:\nrows,4.5);
    \end{axis}
    \end{tikzpicture}
    \end{subfigure}\hfill{}
    \begin{subfigure}{0.48\textwidth}
    \caption{\tool{Secmult-sec-rr}}
    \label{fig:secmult-m4-secure}
    \begin{tikzpicture}[remember picture]
    \pgfplotstablegetrowsof{exps/data_m4/secmult_RR/secmult_twg_e_0.csv}\pgfmathsetmacro\nrows{\pgfmathresult+125}
    \begin{axis}[
        width=6.98cm,
        height=\figheigth,
        table/col sep=comma,
        xlabel={time ($sample$)},
        ylabel={$|t$-test$|$},
        xmin=10, 
        xmax=\nrows,
        ymax=6.5,
        x label style={inner ysep=0pt, outer ysep=0pt},
        title style={inner ysep=0pt, outer ysep=0pt},
        enlargelimits=false,
        axis x line=bottom,
        axis y line=left,
        x tick label style={/pgf/number format/.cd, set thousands separator={}},
        legend style={nodes={scale=0.75, transform shape}},
    ]
        \addplot [Paired-7] table [y expr={abs(\thisrowno{1})}] {exps/data_m4/secmult_RR/secmult_twg_e_0.csv};
        \addlegendentry{Input \texttt{a}}
        \addplot [Paired-1] table [y expr={abs(\thisrowno{1})}] {exps/data_m4/secmult_RR/secmult_twg_e_1.csv};
        \addlegendentry{Input \texttt{b}}
        \draw[Paired-5, dashed, line width=1.2pt] (axis cs:0,4.5) -- (axis cs:\nrows,4.5);
    \end{axis}
    \end{tikzpicture}
    \end{subfigure}
    \caption{$t$-test values for the unsecure (\subref{fig:secmult-m4-unsecure}) and secure (\subref{fig:secmult-m4-secure}) \tool{Secmult} on the Cortex-M4.}
    \label{fig:secmult-m4}
\end{figure}

\subsubsection{In-depth analysis of leakage sources and cycle-accurate comparisons}

We present three cycle-accurate comparisons between \framework{} verdict and $t$-test values on the \tool{Cortex-M4} core: the \tool{DOM-AND-nsec}, \tool{Secmult-nsec}, and an additional \tool{Refresh} micro-benchmark (Figures~\ref{fig:dom-and-m4-aleakator},~\ref{fig:secmult-m4-aleakator} and~\ref{fig:refresh-m4-aleakator} respectively). We additionally show how \framework{}'s output is used to precisely identify micro-architectural leakage sources.

\noindent\textbf{\framework{} verdict and $t$-test traces alignment.}
Four $t$-test values are provided per cycle by our experimental setup. As \framework{} is cycle accurate, only the maximal value of this statistic for the cycle is plotted for the whole cycle. Moreover, as the whole physical system-on-chip may slightly differ from the Verilog model provided to us, we experienced some cycle shifts between the simulated and measured traces. Traces are resynchronised by stuttering items either in the simulation or measured traces. Once expanded, the number of cycles introduced by stuttering is less than 5\% of the overall trace. Results show that all $t$-test values above the 4.5 limit are associated to a leakage detection by \framework{}. We now illustrate for each use-cases some sources of leakage.

\noindent\textbf{DOM-AND-nsec.} For this use-case, each leakage identified with \framework{} is observed using the $t$-test analysis (Figure~\ref{fig:dom-and-m4-aleakator}). We focus on four different sources of leakage, which are numbered along the trace.

\tikzset{saturateA/.style={decorate, decoration={snake,segment length=0.5mm,amplitude=0.5mm,pre=lineto,pre length=0.25mm, post length=0.25mm}}}
\tikzset{saturateB/.style={decorate, decoration={snake,segment length=0.5mm,amplitude=0.5mm,pre=lineto,pre length=0.5mm, post length=0.5mm}}}
\tikzset{saturatelegend/.style={decorate, decoration={snake,segment length=0.5mm,amplitude=0.5mm,pre=lineto,pre length=0.5mm, post length=0.5mm}}}

\setlength{\figheigth}{4cm}
\tikzstyle{pointer} = [draw=black, circle, font=\tiny, inner sep=0.25mm]

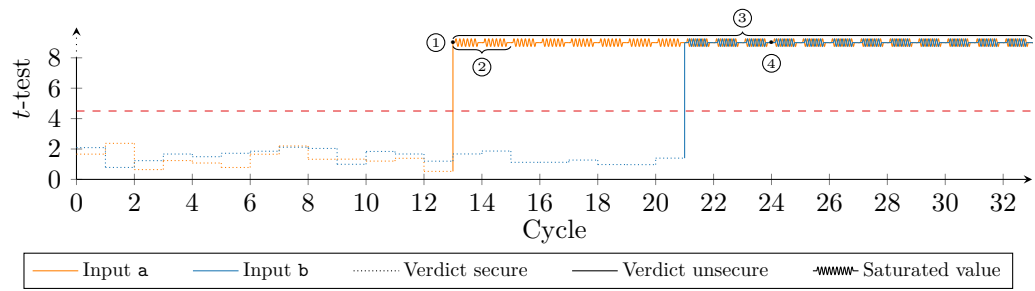
\begin{figure}
    \centering
    \begin{tikzpicture}[remember picture]
    \pgfplotstablegetrowsof{exps/data_m4/dom_and_gprt/cm4-dom-and-gprt-aleakator.csv}\pgfmathsetmacro\finalx{\pgfmathresult-2}
    \pgfplotstableread[col sep=comma]{exps/data_m4/dom_and_gprt/cm4-dom-and-gprt-aleakator.csv}{\first}
    \begin{axis}[
        width=14.15cm,
        height=\figheigth,
        table/col sep=comma,
        xlabel={Cycle},
        ylabel={$t$-test},
        xmin=0,
        xmax=\finalx,
        ymax=12,
        ymin=0,
        x label style={inner ysep=0pt, outer ysep=0pt},
        title style={inner ysep=0pt, outer ysep=0pt},
        enlargelimits=false,
        axis x line=bottom,
        axis y line=left,
        ytick={0,2,...,8},
        y axis line style={draw=none, insert path={(axis cs:0,0) edge[draw, -] (axis cs:0,8) (axis cs:0,8) edge[draw, dotted] (axis cs:0,10)}},
        legend style={
            legend to name=legendaleakatorverdict,
            column sep=15pt,
            legend columns = -1,
            nodes={scale=0.75, transform shape},
            /tikz/every odd column/.append style={column sep=0cm}, 
        },
    ]
        \pgfplotsextra{
            \pgfplotstablegetelem{0}{tA}\of\first
            \let\valyA=\pgfplotsretval
            \pgfplotstablegetelem{0}{tB}\of\first
            \let\valyB=\pgfplotsretval
            \def\prevstyleA{densely dotted}
            \def\prevstyleB{densely dotted}
            \foreach \idx [evaluate=\idx as \nextidx using int(\idx+1)] in {0,...,\finalx} {
                \pgfplotstablegetelem{\nextidx}{vA}\of\first
                \pgfmathsetmacro\styleA{ifthenelse(\pgfplotsretval==0,"densely dotted","solid")}
                
                \pgfplotstablegetelem{\nextidx}{vB}\of\first
                \pgfmathsetmacro\styleB{ifthenelse(\pgfplotsretval==0,"densely dotted","solid")}

                \pgfplotstablegetelem{\nextidx}{tA}\of\first
                \tikzmath{\nextvalyA=min(\pgfplotsretval,9);}
                \pgfplotstablegetelem{\nextidx}{tB}\of\first
                \tikzmath{\nextvalyB=min(\pgfplotsretval,9);}
                \pgfmathsetmacro\supstyleA{ifthenelse(\valyA>=9.0,"saturateA","")}
                \pgfmathsetmacro\supstyleB{ifthenelse(\valyB>=9.0,"saturateB","")}

                \draw[Paired-7,\prevstyleA,\supstyleA] (axis cs:\idx,\valyA) -- (axis cs:\nextidx,\valyA); \draw[Paired-7,\styleA] (axis cs:\nextidx,\valyA) -- (axis cs:\nextidx,\nextvalyA);
                \draw[Paired-1,\prevstyleB,\supstyleB] (axis cs:\idx,\valyB) -- (axis cs:\nextidx,\valyB);
                \draw[Paired-1,\styleB] (axis cs:\nextidx,\valyB) [\styleB] -- (axis cs:\nextidx,\nextvalyB);
                \xdef\valyA{\nextvalyA}
                \xdef\valyB{\nextvalyB}
                \xdef\prevstyleA{\styleA}
                \xdef\prevstyleB{\styleB}
            }
        }
        \addlegendimage{Paired-7}\addlegendentry{Input \texttt{a}}
        \addlegendimage{Paired-1}\addlegendentry{Input \texttt{b}}
        \addlegendimage{densely dotted}\addlegendentry{Verdict secure}
        \addlegendimage{solid}\addlegendentry{Verdict unsecure}
        \addlegendimage{saturatelegend}\addlegendentry{Saturated value}
        \addplot [Paired-5, mark=none, dashed, domain=0:\finalx] {4.5};
        \node[] at (axis cs:13,9) {\Large .};
        \node[left, xshift=-0.1cm, pointer] at (axis cs:13,9) {1};
        \draw [yshift=-0.05cm, decorate, decoration = {brace, mirror}] (axis cs:13,9) -- (axis cs:15,9) node[midway, below, yshift=-0.15cm, pointer]{2};
        \draw [yshift=0.05cm, decorate, decoration = {brace}] (axis cs:13,9) -- (axis cs:33,9) node[midway, above, yshift=0.15cm, pointer]{3};
        \node[] at (axis cs:24,9) {\Large .};
        \node[below, yshift=-0.15cm, pointer] at (axis cs:24,9) {4};
    \end{axis}
    \node[anchor=north] at (current bounding box.south) {\ref*{legendaleakatorverdict}};
    \end{tikzpicture}
    \caption{Cycle's max $t$-test value for each secret and corresponding \framework{} \lmd{RR 1-sw-probing} verdict for the \tool{DOM-AND-nsec} on the Cortex-M4.}
    \label{fig:dom-and-m4-aleakator}
\end{figure}

\circled{1} The first leakage occurs at cycle 13. \framework{} reports the flip-flop signal \texttt{HRDATAS\_RAM} to be leaking in transition without glitches and points to the corresponding HDL description. A further analysis, using program counter and expression evolution on this signal reported by \framework{}, enables to pinpoint the cause of the leakage in the assembly code given in Listing~\ref{lds-dom-and}, whose purpose is to load both shares of the secret input \texttt{a} into the register file.
\begin{lstlisting}[label=lds-dom-and, caption=Load of shares \texttt{a0} and \texttt{a1} in \tool{DOM-AND-nsec}]
ldr r2, =a0  // PC-relative load @a0 in r2
ldr r0, [r2] // a0 in r0
ldr r2, =a1  // PC-relative load @a1 in r2
ldr r1, [r2] // a1 in r1
\end{lstlisting}
The first and third \texttt{ldr} are PC-relative loads, they read the address of the shares \texttt{a0} and \texttt{a1} from the code segment, transiting on the instruction bus from the FLASH memory. The second and fourth \texttt{ldr} read the data in RAM, transiting on the data bus, \texttt{HRDATAS\_RAM}. The third \texttt{ldr} does not clear this bus, hence the transition between both shares is detected.

\circled{2} The transition without glitches reported in \circled{1} at cycle 13 is also reported as a leakage in value with glitches (Register's case of Equation~\ref{eq:lset-def}) and as a leakage in transition with glitches (line \texttt{Expression to verify} in Table~\ref{tb:over-approximation}) at both cycles 13 and 14.

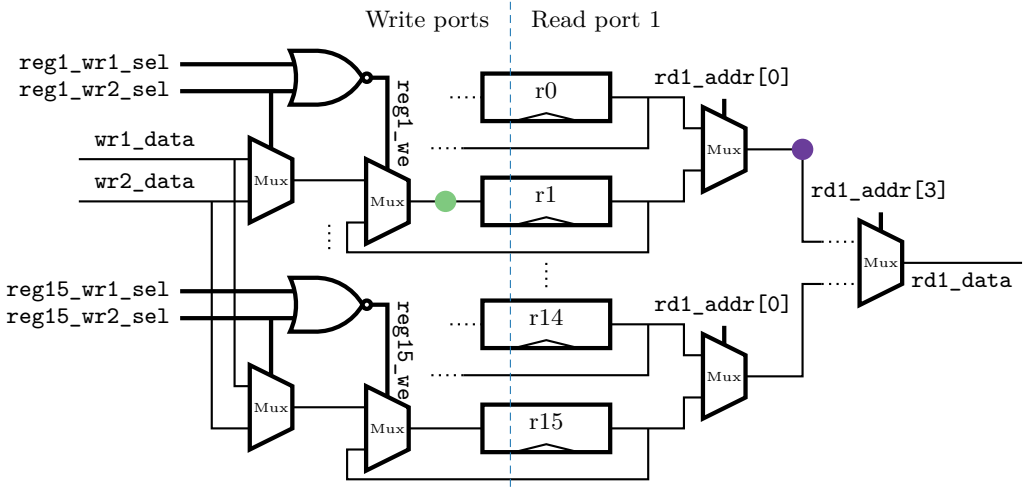
\begin{figure}
    \tikzstyle{mux 2anont} = [muxdemux, muxdemux def={NL=2, NT=1, NB=0, w=1, Rh=1, Lh=2}, no leads]
    \tikzstyle{mux 2anonb} = [muxdemux, muxdemux def={NL=2, NT=0, NB=1, w=1, Rh=1, Lh=2}, no leads]
    \tikzstyle{reg} = [rotate=-90, flipflop, flipflop def={cd=1, t2={\hphantom}, t5={\hphantom}, clock wedge size=0.5}, scale=0.5, external pins width=0]
    \tikzstyle{cpureg} = [flipflop, flipflop def={cd=1, t2={\hphantom}, t5={\hphantom}, clock wedge size=0.5}, scale=0.5, xscale=2, yscale=0.5, external pins width=0]
    \tikzstyle{llabel} = [text=lset]
    \tikzstyle{elabel} = [text=symb]
    \tikzstyle{stable} = [draw=isstab]
    \tikzstyle{unstable} = [draw=isunstab]
    \begin{circuitikz}[node distance=0.5cm]
        \node (x1) [cpureg, thick] {};
        \node (x2) [cpureg, thick, below=0.75cm of x1] {};
        \node (x3) [cpureg, thick, below=1.00cm of x2] {};
        \node (x4) [cpureg, thick, below=0.75cm of x3] {};
        \node (x5) [thick, below=0.25cm of x4] {};

        \node[mux 2anont, thick] (muxw1) at ([xshift=-1.25cm]x2.pin 2) {\tiny Mux};
        \node[mux 2anont, thick] (muxw2) at ([xshift=-1.25cm]muxw1.blpin 1) {\tiny Mux};
        \node[mux 2anont, thick] (muxw3) at ([xshift=-1.25cm]x4.pin 2) {\tiny Mux};
        \node[mux 2anont, thick] (muxw4) at ([xshift=-1.25cm]muxw3.blpin 1) {\tiny Mux};

        \node[mux 2anont, thick] (muxr1) at ([xshift=1.5cm]$(x1.pin 5)!0.5!(x2.pin 5)$) {\tiny Mux};
        \node[mux 2anont, thick] (muxr2) at ([xshift=1.5cm]$(x3.pin 5)!0.5!(x4.pin 5)$) {\tiny Mux};
        \node[mux 2anont, thick] (muxr3) at ([xshift=1.5cm]$(muxr1.rpin 1)!0.5!(muxr2.rpin 1)$) {\tiny Mux};

        \node (rpa) [above=0cm of muxr3.brpin 1, anchor=north west] {\small \texttt{rd1\_data}};

        \node (lt1) [above=0cm of muxr1.tpin 1, text width=2cm, align=center] {\small \texttt{rd1\_addr[0]}};
        \node (lt2) [above=0cm of muxr2.tpin 1, text width=2cm, align=center] {\small \texttt{rd1\_addr[0]}};
        \node (lt3) [above=0cm of muxr3.tpin 1, text width=2cm, align=center] {\small \texttt{rd1\_addr[3]}};

        \node (mw1) [left=0.5cm of muxw2.blpin 1, anchor=south east, text width=1.5cm, align=center] {\small \texttt{wr1\_data}};
        \node (mw2) [left=0.5cm of muxw2.blpin 2, anchor=south east, text width=1.5cm, align=center] {\small \texttt{wr2\_data}};

        \node[nor port, thick, scale=0.85, no leads] (orw1) at ([yshift=1.5cm]$(muxw2.east)!0.5!(muxw1.west)$) {};
        \node[nor port, thick, scale=0.85, no leads] (orw2) at ([yshift=1.5cm]$(muxw4.east)!0.5!(muxw3.west)$) {};

        \node (mw1s1) [left=1.5cm of orw1.bin 1] {\small \texttt{reg1\_wr1\_sel}};
        \node (mw1s2) [left=1.5cm of orw1.bin 2] {\small \texttt{reg1\_wr2\_sel}};
                
        \node (mw3s1) [left=1.5cm of orw2.bin 1] {\small \texttt{reg15\_wr1\_sel}};
        \node (mw3s2) [left=1.5cm of orw2.bin 2] {\small \texttt{reg15\_wr2\_sel}};

        \node (lx1) [below=0cm of x1.north] {\small r0};
        \node (lx2) [below=0cm of x2.north] {\small r1};
        \node (lx3) [below=0cm of x3.north] {\small r14};
        \node (lx3) [below=0cm of x4.north] {\small r15};
        \draw [thick, dotted] (x2.south)+(0, -0.5cm) -- +(0, -0.85cm);
        
        \draw [thick, dotted] ($(muxw2.east)!0.5!(muxw1.west)$)+(0, -0.45cm) -- +(0, -0.8cm);

        \draw [thick] (x1.pin 5) -| ([xshift=-0.25cm]muxr1.blpin 1) -- (muxr1.blpin 1);
        \draw [thick] (x2.pin 5) -| ([xshift=-0.25cm]muxr1.blpin 2) -- (muxr1.blpin 2);
        \draw [thick] (x3.pin 5) -| ([xshift=-0.25cm]muxr2.blpin 1) -- (muxr2.blpin 1);
        \draw [thick] (x4.pin 5) -| ([xshift=-0.25cm]muxr2.blpin 2) -- (muxr2.blpin 2);
        
        \draw [thick] (muxr1.brpin 1) -- ([xshift=0.75cm]muxr1.brpin 1) node [fill=Paired-9, circle, inner sep=0.1cm] {} |- ([xshift=-0.5cm]muxr3.blpin 1);
        \draw [thick] (muxr2.brpin 1) -- ([xshift=0.75cm]muxr2.brpin 1) |- ([xshift=-0.5cm]muxr3.blpin 2);
        \draw [thick, dotted] ([xshift=-0.5cm]muxr3.blpin 1) -- (muxr3.blpin 1);
        \draw [thick, dotted] ([xshift=-0.5cm]muxr3.blpin 2) -- (muxr3.blpin 2);

        \draw [thick] (muxr3.brpin 1) -- (rpa.north east);
        \draw [ultra thick] ([yshift=0.25cm]muxr1.btpin 1) -- (muxr1.btpin 1);
        \draw [ultra thick] ([yshift=0.25cm]muxr2.btpin 1) -- (muxr2.btpin 1);
        \draw [ultra thick] ([yshift=0.25cm]muxr3.btpin 1) -- (muxr3.btpin 1);

        \draw [thick] ([xshift=0.5cm]x1.pin 5) -- ([xshift=0.5cm, yshift=-0.675cm]x1.pin 5) -- ([xshift=-0.25cm, yshift=-0.675cm]x1.pin 5-|x1.pin 2);
        \draw[thick, dotted] ([xshift=-0.25cm, yshift=-0.675cm]x1.pin 5-|x1.pin 2) -- +(-0.45cm, 0);
        \draw[thick, dotted] (x1.pin 2) -- +(-0.45cm, 0);

        \draw [thick] ([xshift=0.5cm]x2.pin 5) -- ([xshift=0.5cm, yshift=-0.675cm]x2.pin 5) -- ([xshift=-0.25cm, yshift=-0.675cm]x2.pin 5-|muxw1.blpin 2) |- (muxw1.blpin 2);
        \draw[thick] (muxw1.brpin 1) -- (x2.pin 2) node [midway, fill=Accent-1, circle, inner sep=0.1cm] {};

        \draw [thick] ([xshift=0.5cm]x3.pin 5) -- ([xshift=0.5cm, yshift=-0.675cm]x3.pin 5) -- ([xshift=-0.25cm, yshift=-0.675cm]x3.pin 5-|x3.pin 2);
        \draw[thick, dotted] ([xshift=-0.25cm, yshift=-0.675cm]x3.pin 5-|x3.pin 2) -- +(-0.45cm, 0);
        \draw[thick, dotted] (x3.pin 2) -- +(-0.45cm, 0);
        
        \draw [thick] ([xshift=0.5cm]x4.pin 5) -- ([xshift=0.5cm, yshift=-0.675cm]x4.pin 5) -- ([xshift=-0.25cm, yshift=-0.675cm]x4.pin 5-|muxw3.blpin 2) |- (muxw3.blpin 2);
        \draw[thick] (muxw3.brpin 1) -- (x4.pin 2);

        \draw [thick] (muxw1.blpin 1) -- (muxw2.brpin 1);
        \draw [thick] (muxw3.blpin 1) -- (muxw4.brpin 1);

        \draw [thick] (muxw2.blpin 1) -- (muxw2.blpin 1-|mw1.south west);
        \draw [thick] (muxw2.blpin 2) -- (muxw2.blpin 2-|mw2.south west);
        
        \draw [thick] ([xshift=-0.2cm]muxw2.blpin 1) |- (muxw4.blpin 1);
        \draw [thick] ([xshift=-0.5cm]muxw2.blpin 2) |- (muxw4.blpin 2);

        \draw[ultra thick] (orw1.bout) -| (muxw1.btpin 1) node (lw1) [near end, xshift=0.2cm, rotate=-90, text width=2cm, align=center] {\small \texttt{reg1\_we}};
        \draw[ultra thick] (mw1s2.east) -- (orw1.bin 2);
        \draw[ultra thick] (mw1s1.east) -- (orw1.bin 1);
        \draw[ultra thick] (muxw2.btpin 1|-orw1.bin 2) -- (muxw2.btpin 1);
        
        \draw[ultra thick] (orw2.bout) -| (muxw3.btpin 1) node (lw2) [near end, xshift=0.2cm, rotate=-90, text width=2cm, align=center] {\small \texttt{reg15\_we}};
        \draw[ultra thick] (mw3s2.east) -- (orw2.bin 2);
        \draw[ultra thick] (mw3s1.east) -- (orw2.bin 1);
        \draw[ultra thick] (muxw4.btpin 1|-orw2.bin 2) -- (muxw4.btpin 1);

        \draw[densely dashed, Paired-1] (current bounding box.north)+(0, 0.5cm) -- (current bounding box.south);
        \node[below right=0cm and 0.15cm of current bounding box.north] {\small Read port 1};
        \node[below left=0cm and 0.15cm of current bounding box.north] {\small Write ports};
    \end{circuitikz}
    \caption{ Simplified view of the register file of the Cortex-M4 core.}
    \label{fig:regfile}
\end{figure}

\begin{figure}
    \centering
    \begin{tikzpicture}[remember picture]
    \pgfplotstablegetrowsof{exps/data_m4/refresh/refresh.csv}\pgfmathsetmacro\finalx{\pgfmathresult-2}
    \pgfplotstableread[col sep=comma]{exps/data_m4/refresh/refresh.csv}{\first}
    \begin{axis}[
        width=14.15cm,
        height=\figheigth,
        table/col sep=comma,
        xlabel={Cycle},
        ylabel={$t$-test},
        xmin=0,
        xmax=\finalx,
        ymax=12,
        ymin=0,
        x label style={inner ysep=0pt, outer ysep=0pt},
        title style={inner ysep=0pt, outer ysep=0pt},
        enlargelimits=false,
        axis x line=bottom,
        axis y line=left,
        y axis line style={draw=none, insert path={(axis cs:0,0) edge[draw, -] (axis cs:0,8) (axis cs:0,8) edge[draw, dotted] (axis cs:0,10)}},
        legend style={
            legend to name=legendaleakatorverdict2,
            column sep=15pt,
            legend columns = -1,
            nodes={scale=0.75, transform shape},
            /tikz/every odd column/.append style={column sep=0cm}, 
        },
        ytick={0,2,...,8},
    ]
        \pgfplotsextra{
            \pgfplotstablegetelem{0}{tA}\of\first
            \let\valyA=\pgfplotsretval
            \def\prevstyleA{densely dotted}
            \foreach \idx [evaluate=\idx as \nextidx using int(\idx+1)] in {0,...,\finalx} {
                \pgfplotstablegetelem{\nextidx}{vA}\of\first
                \pgfmathsetmacro\styleA{ifthenelse(\pgfplotsretval==0,"densely dotted","solid")}

                \pgfplotstablegetelem{\nextidx}{tA}\of\first
                \tikzmath{\nextvalyA=min(\pgfplotsretval,9);}
                \pgfmathsetmacro\supstyleA{ifthenelse(\valyA>=9.0,"saturateA","")}

                \draw[Paired-7,\prevstyleA,\supstyleA] (axis cs:\idx,\valyA) -- (axis cs:\nextidx,\valyA); \draw[Paired-7,\styleA] (axis cs:\nextidx,\valyA) -- (axis cs:\nextidx,\nextvalyA);
                \xdef\valyA{\nextvalyA}
                \xdef\prevstyleA{\styleA}
            }
        }
        \addlegendimage{Paired-7}\addlegendentry{Input \texttt{a}}
        \addlegendimage{densely dotted}\addlegendentry{Verdict secure}
        \addlegendimage{solid}\addlegendentry{Verdict unsecure}
        \addlegendimage{saturatelegend}\addlegendentry{Saturated value}
        \addplot [Paired-5, mark=none, dashed, domain=0:\finalx] {4.5};
    \end{axis}
    \node[anchor=north] at (current bounding box.south) {\ref*{legendaleakatorverdict2}};
    \end{tikzpicture}
    \caption{Cycle's max $t$-test value for each secret and corresponding \framework{} \lmd{RR 1-sw-probing} verdict for the \tool{Refresh} on the Cortex-M4.}
    \label{fig:refresh-m4-aleakator}
\end{figure}
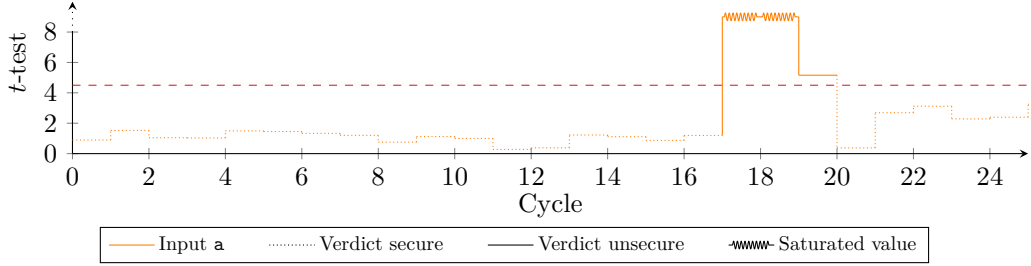

\circled{3} \framework{} reports a leakage by value with glitches from cycle 13 --- performing the write of the second share of input \texttt{a} in the register file --- onwards. This leakage is linked with the signal \texttt{rd1\_data} of the \texttt{u\_cm4\_dpu\_regbank} module, which is the output of the multiplexer tree corresponding to the read port 1 of the register file. An analysis of the \lss{} associated to this reported leakage shows that both shares are simultaneously present in the register file and may be recombined due to glitches. Part of read port 1 is pictured in Figure~\ref{fig:regfile}: if the multiplexer selector driven by the \texttt{rd1\_addr[0]} wire is unstable, the multiplexer tree may glitch the \texttt{r0} and \texttt{r1} registers content. This explains why the \ls{} of the wire marked as \tikz[baseline=(char.south)] \node [midway, fill=Paired-9, circle, inner sep=0.075cm] (char) {}; contains both the \lss{} of \texttt{r0} ($\{$\texttt{a0}$\}$) and \texttt{r1} ($\{$\texttt{a1}$\}$), leading to a possible observation of the secret \texttt{a}.  

\circled{4} At cycle 24, \framework{} identifies a leakage related to the register file and the execution of the \texttt{xor} instruction \lstinline[label=write-reg-leak]$eor r4, r4, r3 // r4 = (a0 and b1), r3 = m$.
The identified leakage source is the flip-flop holding \texttt{r1} value in the register file. The design of the write path in the register file, as illustrated in Figure~\ref{fig:regfile}, may provoke glitches between the written data and the content of each register when its write enable signal  is unstable. When writing the result of the \texttt{eor} instruction of \tool{DOM-AND-nsec}, the \texttt{reg1\_we} is unstable. As a consequence, the \ls{} of the wire marked as \tikz[baseline=(char.south)] \node [midway, fill=Accent-1, circle, inner sep=0.075cm] (char) {}; contains both the \lss{} of the write ports \texttt{wr1\_data}  and \texttt{wr2\_data} as well as the \ls{} of the \texttt{r1} register.
The \ls{} of the write port \texttt{wr1\_data} is equal to the \ls{} of the ALU output, i.e. $\{a0 \& b1, m\}$. The \ls{} of the \texttt{r1} register is $\{a1\}$. As a consequence the \ls{} of the wire marked as \tikz[baseline=(char.south)] \node [midway, fill=Accent-1, circle, inner sep=0.075cm] (char) {}; may leak \texttt{a}.

Similar phenomenons explain the leakages related to Input \texttt{b}, from the load of its shares at cycle 21 onwards.

\noindent\textbf{Refresh micro-benchmark.} \framework{} can be used to identify leakages in seemingly non-leaking programs. Listing~\ref{refresh} details a program loading a share \texttt{a0} from a secret \texttt{a}, refreshing it with a mask \texttt{m} and storing it. Then, after clearing the register, the second share \texttt{a1} is loaded, refreshed with the same mask and stored.

The Armv7-M architecture used in the Cortex-M4 contains both 16-bit and 32-bit encoded instructions. Most instructions have multiple encodings while ensuring the same functional behaviour. For example, the \texttt{mov} instruction has three encoding, including two of 32-bit while the \texttt{ldr} instruction has two encodings, one 32-bit and one 16-bit. The micro-architectural behaviour of the instructions is dependent of the encoding~\cite{armistice}. In the case of both the \texttt{mov} and \texttt{ldr} instructions, the written ALU ports differ with their encodings.

Since the LSU read and write data paths, along with the ALU and registers, are cleared between the loads, one could expect this implementation to be secure. However, as \texttt{r4} is cleared by \texttt{xor}ing it with itself, its value is written in both read ports of the ALU. The following \texttt{mov} is encoded as a 32-bit instruction which only clears the input port \texttt{A} of the ALU.
After this, the 16-bit encoded \texttt{ldr} is executed and also only overwrites the input port \texttt{A} of the ALU. Finally, when the \tool{xor} performing the refresh of \texttt{a1} is executed, input port \texttt{A} transitions from $0$ to $a1$ and input port \texttt{B} transitions from $a0 \oplus m$ to $m$. Taken independently, none of these transitions can leak the secret \texttt{a}, but they cause a leaking transition in internal wires of the ALU (in the multi-cycles operators). This transition without glitches appears at cycle 18 in Figure~\ref{fig:refresh-m4-aleakator}. \framework{} reports leakages in transitions with glitches in the cycles before and after, as the port \tool{A} of the ALU can be recombined with the share \texttt{a1} that is newly loaded before the \texttt{xor}.

\begin{lstlisting}[multicols=2, label=refresh, caption=Refresh program]
ldr r0, =a0
ldr r1, =a1
ldr r2, =m
ldr r3, =blk
ldr r4, [r0] // a0 in r4
ldr r5, [r2] // m in r5

eor r4, r4, r5 // Refresh a0
str r4, [r0]

// Clear write LSU path
str r3, [r3]
eor r4, r4, r4 // Clear r4
mov r4, 0x0 // Clear alu path?

// LSU Read path is cleared
ldr r6, [r1] // a1 in r12

eor r6, r6, r5 // Refresh a1
str r6, [r2]
mov r6, 0x0
\end{lstlisting}
%

These subtle leakages can be prevented by either clearing \texttt{r4} without rewriting it on the ALU ports with the \texttt{xor} (e.g \lstinline$eor r4, r13, r13$), or using the 16-bit encoding of the \texttt{mov}, or using the 32-bit encoding of the \texttt{ldr}.

This real-world example is a key-argument of using automated tools such as \framework{}: clearing data paths is difficult, it requires advanced knowledge of the data path taken by all instruction encodings and requires to take into account all the effects of surrounding instructions.

\noindent\textbf{Secmult-nsec.} Figure~\ref{fig:secmult-m4-aleakator} demonstrates that all cycles for which a leakage is observed in practice are correctly identified as leaking by \framework{}. Since this program is compiled with \texttt{-O3} optimisations, the masking in the source code is not perfectly preserved throughout the entire program. We also note that there are cycles without any observable leakages which are still reported as leaking by the tool, and we identified four main reasons for this discrepancy.

Firstly, an analysis of the compiled program shows that the four shares of \texttt{a} and \texttt{b} are stored within the same 32-bit memory word in the \lstinline$.data$ section. As a result, \framework{} logically reports all memory reads of input shares as leaking in value without glitches (cycles 4-18 in Figure~\ref{fig:secmult-m4-aleakator}); even for byte loads, 32-bit words of data are put on the bus and later realigned and filtered in the LSU. Still, no leakage is observed in the $t$-test measurements, which we attribute to two reasons: 1) the $t$-test is performed at first order while having two shares in a single word is a second order leakage, and 2) from the point of view of a single secret variable, other shares in the same word behaves as a noise, which can be sufficient to hide the leakage.

Second, the analysis of the remaining detected but non-observable leakages shows that most are related to the presence of all shares in the register file, potentially leaking with glitches as illustrated in Figure~\ref{fig:regfile}. This may be linked to a third reason for the mismatch: the modelling of glitches is an over-approximation of real glitches, and only a subset of them may actually manifest in practice, depending on the synthesis timings.

Finally, the fourth reason for the mismatch between predicted and observed leakages is that we cannot guarantee that the verified RTL is exactly the same as the one used to produce the actual chip.

\tikzset{saturateA/.style={decorate, decoration={snake,segment length=0.3mm,amplitude=0.5mm,pre=lineto,pre length=0.0mm, post length=0.25mm}}}
\tikzset{saturateB/.style={decorate, decoration={snake,segment length=0.3mm,amplitude=0.5mm,pre=lineto,pre length=0.25mm, post length=0.0mm}}}

\begin{figure}
    \centering
    \begin{tikzpicture}[remember picture]
    \pgfplotstablegetrowsof{exps/data_m4/secmult_unsecure/cm4-secmult-aleakator.csv}\pgfmathsetmacro\finalx{\pgfmathresult-2}
    \pgfplotstableread[col sep=comma]{exps/data_m4/secmult_unsecure/cm4-secmult-aleakator.csv}{\first}
    \begin{axis}[
        width=14.15cm,
        height=\figheigth,
        table/col sep=comma,
        xlabel={Cycle},
        ylabel={$t$-test},
        xmin=0,
        xmax=\finalx,
        ymax=9.25,
        ymin=0,
        x label style={inner ysep=0pt, outer ysep=0pt},
        title style={inner ysep=0pt, outer ysep=0pt},
        enlargelimits=false,
        axis x line=bottom,
        axis y line=left,
        ytick={0,2,...,8},
        y axis line style={draw=none, insert path={(axis cs:0,0) edge[draw, -] (axis cs:0,8) (axis cs:0,8) edge[draw, dotted] (axis cs:0,10)}},
        legend style={nodes={scale=0.75, transform shape}},
    ]
        \addplot [Paired-5, mark=none, dashed, domain=0:\finalx] {4.5};
        \pgfplotsextra{
            \pgfplotstablegetelem{0}{tA}\of\first
            \let\valyA=\pgfplotsretval
            \pgfplotstablegetelem{0}{tB}\of\first
            \let\valyB=\pgfplotsretval
            \def\prevstyleA{densely dotted}
            \def\prevstyleB{densely dotted}
            \foreach \idx [evaluate=\idx as \nextidx using int(\idx+1)] in {0,...,\finalx} {
                \pgfplotstablegetelem{\nextidx}{vA}\of\first
                \pgfmathsetmacro\styleA{ifthenelse(\pgfplotsretval==0,"densely dotted","solid")}
                
                \pgfplotstablegetelem{\nextidx}{vB}\of\first
                \pgfmathsetmacro\styleB{ifthenelse(\pgfplotsretval==0,"densely dotted","solid")}

                \pgfplotstablegetelem{\nextidx}{tA}\of\first
                \tikzmath{\nextvalyA=min(\pgfplotsretval,9);}
                \pgfplotstablegetelem{\nextidx}{tB}\of\first
                \tikzmath{\nextvalyB=min(\pgfplotsretval,9);}
                \pgfmathsetmacro\supstyleA{ifthenelse(\valyA>=9.0,"saturateA","")}
                \pgfmathsetmacro\supstyleB{ifthenelse(\valyB>=9.0,"saturateB","")}

                \draw[Paired-7,\prevstyleA,\supstyleA] (axis cs:\idx,\valyA) -- (axis cs:\nextidx,\valyA); \draw[Paired-7,\styleA] (axis cs:\nextidx,\valyA) -- (axis cs:\nextidx,\nextvalyA);
                \draw[Paired-1,\prevstyleB,\supstyleB] (axis cs:\idx,\valyB) -- (axis cs:\nextidx,\valyB);
                \draw[Paired-1,\styleB] (axis cs:\nextidx,\valyB) [\styleB] -- (axis cs:\nextidx,\nextvalyB);
                \xdef\valyA{\nextvalyA}
                \xdef\valyB{\nextvalyB}
                \xdef\prevstyleA{\styleA}
                \xdef\prevstyleB{\styleB}
            }
        }
    \end{axis}
    \node[anchor=north] at (current bounding box.south) {\ref*{legendaleakatorverdict}};
    \end{tikzpicture}
    \caption{Cycle's max $t$-test value for each secret and corresponding \framework{} \lmd{RR 1-sw-probing} model verdict for the \tool{Secmult-nsec} on the Cortex-M4.}
    \label{fig:secmult-m4-aleakator}
\end{figure}
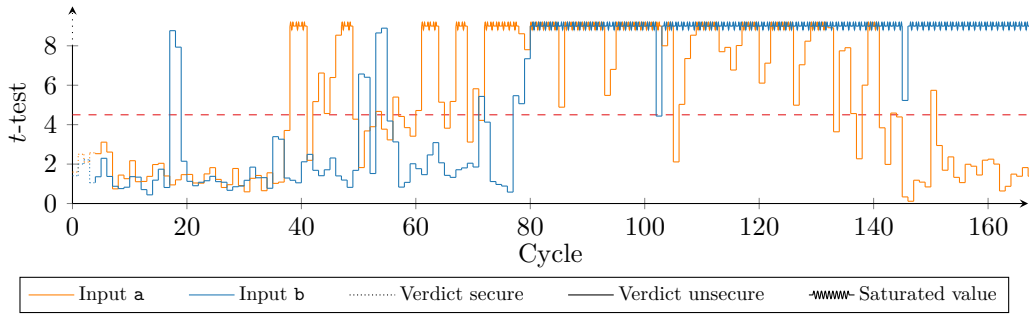

As a conclusion for this cycle-accurate analysis pointing to the source of leakages, we want to recall that all observed leakages are reported by \framework{}. Thanks to the precise output of \framework{}, all the reported leakages have been circumvented by introducing barriers, data variables reorganisation and shuffling instructions where needed, leading to the secure versions \tool{DOM-AND-sec} and \tool{Secmult-sec}.

\subsection{Higher-order and Other Security Properties}

\noindent\textbf{Higher-order verification.} Multiple propositions have been made regarding how to define higher-order verification for hardware and software. There can be spatial combinations, temporal combinations or a mix of both ~\cite{CocoAlma,TCHES:MulMor22}. The common point of these propositions is that a verification at order $d$ requires the analysis of $d$-uplets of expressions or \lss{}, which must be verified together. \framework{} is able to handle both spatial and temporal combinations, as all the valuations required for such a verification are already computed at each cycle. However, no optimization are currently implemented, e.g. no optimisations regarding redundant probes. Therefore, \framework{} is not yet meant to be used for higher-order verification on large circuits. Besides, even with optimisations, all existing higher-order verification methods suffer scalability issues because of the combinatorial increase in the number of $d$-uplets to verify. The latter grows as the number of combinations of $d$ elements among $p$, with $p$ being respectively the number of wires\footnote{or bits for single bits leakage models}, simulated cycles, or the product of both, for spatial, temporal and mixed verification.

To support this discussion, we present in Table~\ref{tb:ho-res} the verdict of \framework{} on the analysis of two hardware accelerators at different orders, reproducing existing spatial verifications on \tool{DOM-KECCAK}~\cite{keccakHO} (\coco{} and \prover{}) and \tool{DOM-AND}~\cite{grossDomainOrientedMaskingCompact2016} (\tool{VerifMSI}). The verdicts of \framework{} are the expected ones. Moreover, results show the explosion in the number of $d$-uplets to enumerate and verify as the order grows; it quickly becomes intractable, as, for example, the verification at order 3 of a 4-share \tool{DOM-KECCAK} would require the enumeration of more than 34 billions triplets.

\begin{table}
    \small
    \centering
    \begin{tabular}{l|r|r|r|c|r}
        \hline
        \makecell{Masked} & \multirow{2}{*}{Cycles} & \makecell{Security} & \makecell{Verification} & \makecell{\framework{}} & \makecell{$\sharp$ $d$-uplets} \\
        \makecell{hardware} & & \makecell{Order} & \makecell{Order $d$} & \makecell{verdict} & \makecell{to Verify} \\ \hline
        \tool{DOM-AND} & 2/2 & 1 & 2 & \xmark & 240 \\
        \tool{DOM-AND} & 2/2 & 2 & 2 & \cmark & 666 \\
        \tool{DOM-AND} & 2/2 & 2 & 3 & \xmark & 15 540 \\
        \tool{DOM-AND} & 2/2 & 5 & 4 & \cmark & 450 662 52 \\ \hline
        \tool{DOM-KECCAK} & 2/2 & 1 & 1 & \cmark & 918 \\
        \tool{DOM-KECCAK} & 2/2 & 2 & 2 & \cmark & 1 047 552 \\
        \tool{DOM-KECCAK} & 2/2 & 2 & 3 & \xmark & 356 866 048 \\
        \tool{DOM-KECCAK} & 2/2 & 3 & 2 & \cmark & 22 028 942 \\
        \hline
    \end{tabular}
    \caption{Higher-order verification results obtained by \framework{}.}
    \label{tb:ho-res}
\end{table}

\noindent\textbf{Other security properties.} \framework{} main purpose is the verification of \lmd{d-probing} properties for software implementations. However, it can verify other security properties for hardware accelerators, such as NI and SNI~\cite{EC:BBDFGS15,CCS:BBDFGS16}, which are both currently implemented. As long as the needed information can be extracted from expressions or sets of expressions, there are no intrinsic limitation to support other security properties such as PINI~\cite{cassiers2020trivially}.

In order to highlight this aspect, we show in Table~\ref{tb:ni_sni_results} the results of NI and SNI verifications of common hardware gadgets \tool{ISW-AND}~\cite{C:IshSahWag03} and \tool{DOM-AND}~\cite{grossDomainOrientedMaskingCompact2016} at order 2 with and without glitches. The \tool{DOM-AND} for this experiment includes additional registers to account for the propagation on multiple cycles. These results corroborate with existing verifications from \tool{VerifMSI}~\cite{verifmsi}.

\begin{table}
    \small
    \centering
    \begin{tabular}{l|r|r|r|c|c|c}
        \hline
        \makecell{Masked} & \multirow{2}{*}{Cycles} & \makecell{Security} & \makecell{Verification} & \multirow{2}{*}{\makecell{Property}} & \makecell{With} & \makecell{$\sharp$ $d$-uplets} \\
        \makecell{hardware} & & \makecell{Order} & \makecell{Order $d$} & & \makecell{glitches} & \makecell{to Verify} \\ \hline
        \tool{DOM-AND} & 2/2 & 2 & 2 & NI & \xmark & \cmark \\
        \tool{DOM-AND} & 2/2 & 2 & 2 & NI & \cmark & \cmark \\
        \tool{DOM-AND} & 2/2 & 2 & 2 & SNI & \xmark & \cmark \\
        \tool{DOM-AND} & 2/2 & 2 & 2 & SNI & \cmark & \xmark \\ \hline
        \tool{ISW-AND} & 2/2 & 2 & 2 & NI & \xmark & \cmark \\
        \tool{ISW-AND} & 2/2 & 2 & 2 & NI & \cmark & \xmark \\
        \tool{ISW-AND} & 2/2 & 2 & 2 & SNI & \xmark & \cmark \\
        \tool{ISW-AND} & 2/2 & 2 & 2 & SNI & \cmark & \xmark \\
        \hline
    \end{tabular}
    \caption{NI and SNI results for hardware gadgets \tool{DOM-AND} and \tool{ISW-AND}}
    \label{tb:ni_sni_results}
\end{table}

\subsection{Discussion}

When using \framework{} on various circuits, we identified discussion-worthy elements, which we detail in this section.

\noindent\textbf{Concretisations.} Verified applications are assumed to have a control flow independent of input variables, represented as symbolic variables (secrets, masks and public variable). During the mixed-simulation, this translates into the constraint that all control wires should have a symbolic expression equivalent to a \token{CST} and their \lss{} should be $\varnothing$. However, this constraint may not hold in some  processors, e.g. due to micro-architectural recombinations, for example when computing flags or bypasses. In such cases, \framework{} propagates, on the control flow gates, non-constant symbolic expressions and non-empty \lss{}, expecting that they will be overridden before being stored in the CPU state. The worst case would be to obtain either a symbolic PC value or a symbolic instruction that would prevent from further simulating the circuit. In all our experiments, our assumption is sufficient and enables the simulation of complex programs on various CPUs.

\noindent\textbf{Scalability.} As for other formal approaches, not all circuits can be fully simulated. The data structure used for verification can cause memory issues when expressions are deep and involve many masks. The cost and simulability highly depend on the circuit structure and the generated expressions. This can be observed in Section~\ref{sc:results} where some small hardware accelerators are heavier to simulate than full CPUs running programs.

\noindent\textbf{False positives.} Our method is not exempt of false positive leaking verdicts as we report an issue when the verification carried out by \tool{VerifMSI++} finds a secret leakage or cannot conclude about the absence of secret leakage. Other formal tools suffer from this issue, as shown by \prover{}~\cite{prover} with \coco{}.


\noindent\textbf{No prior knowledge on CPU.} \framework{} was used on five CPUs without requiring prior knowledge on their implementation. The precision of the output of \framework{} is valuable for understanding the leaks and designing leak-free programs.

\noindent\textbf{Benefits of using expressions.} Using symbolic expressions for both verification and hardware simulation offers many advantages. First, it allows the computation of stability by checking the equivalence of symbolic valuations across two consecutive cycles, without being limited to wires with constant values. Second, it enables the integration of custom rules for memory accesses, such as reading from a table with a symbolic offset. Third, it makes it possible to fall back to enumeration when verification cannot conclude for a small expression, helping to avoid false positives. Finally, the value of a signal can be made concrete at any time using the concrete input values available during simulation. This is useful for checking both the consistency of the expression and the symbolic state.

None of these advantages are possible with \coco{}, which only computes so-called correlation sets. These sets track correlations between the secret inputs and their combinations for each element in the circuit, but do not retain full symbolic information.

\tikzset{saturateA/.style={decorate, decoration={snake,segment length=0.3mm,amplitude=0.5mm,pre=lineto,pre length=0.0mm, post length=0.25mm}}}
\tikzset{saturateB/.style={decorate, decoration={snake,segment length=0.3mm,amplitude=0.5mm,pre=lineto,pre length=0.25mm, post length=0.0mm}}}

\section{Conclusion}
\label{sc:conclusion}

We presented a new method, implemented in a tool called \framework{}, to formally verify cryptographic masked implementations running on CPUs as well as cryptographic hardware accelerators. This method relies on mixed-domain simulation, for which we formally defined the behaviour, promoting clarity in the verifications performed and enabling reproducibility of results. Additionally, both the implementation and the benchmarks are open source. While previous works were limited to formally verifying small programs such as S-boxes on a single CPU, we demonstrated the utility and efficiency of \framework{} by verifying complete real-world software on multiple CPUs.

Furthermore, we showed---using actual power measurements on two CPU cores---that the verification results from \framework{} align with the secret leakages observed in practice. We then presented a cycle accurate comparison between the output of \framework{} and the $t$-test for multiple 
programs on the Cortex-M4 while providing precise and insightful leakage analyses. These verifications required only the HDL description of the circuits, with no additional knowledge about their implementation. Our method scales to real-world CPUs running widely-used cryptographic software, though this may require high RAM consumption. Nonetheless, proofs performed in the \lmd{RR 1-sw-probing} model can quickly guarantee the absence of first-order secret leakages.

\bibliographystyle{alpha}
\bibliography{abbrev3,crypto,custom}

\newcommand{\etalchar}[1]{$^{#1}$}
\begin{thebibliography}{BEOMHE19}

\bibitem[BBC{\etalchar{+}}19]{ESORICS:BBCFGS19}
Gilles Barthe, Sonia Bela{\"i}d, Ga{\"e}tan Cassiers, Pierre-Alain Fouque,
  Benjamin Gr{\'e}goire, and Fran{\c c}ois-Xavier Standaert.
\newblock {maskVerif}: Automated verification of higher-order masking in
  presence of physical defaults.
\newblock In Kazue Sako, Steve Schneider, and Peter Y.~A. Ryan, editors, {\em
  ESORICS~2019, Part~I}, volume 11735 of {\em {LNCS}}, pages 300--318.
  Springer, Cham, September 2019.

\bibitem[BBD{\etalchar{+}}15]{EC:BBDFGS15}
Gilles Barthe, Sonia Bela{\"i}d, Fran{\c c}ois Dupressoir, Pierre-Alain Fouque,
  Benjamin Gr{\'e}goire, and Pierre-Yves Strub.
\newblock Verified proofs of higher-order masking.
\newblock In Elisabeth Oswald and Marc Fischlin, editors, {\em EUROCRYPT~2015,
  Part~I}, volume 9056 of {\em {LNCS}}, pages 457--485. Springer, Berlin,
  Heidelberg, April 2015.

\bibitem[BBD{\etalchar{+}}16]{CCS:BBDFGS16}
Gilles Barthe, Sonia Bela{\"i}d, Fran{\c c}ois Dupressoir, Pierre-Alain Fouque,
  Benjamin Gr{\'e}goire, Pierre-Yves Strub, and R{\'e}becca Zucchini.
\newblock Strong non-interference and type-directed higher-order masking.
\newblock In Edgar~R. Weippl, Stefan Katzenbeisser, Christopher Kruegel,
  Andrew~C. Myers, and Shai Halevi, editors, {\em ACM CCS 2016}, pages
  116--129. {ACM} Press, October 2016.

\bibitem[BEOMHE19]{jcenines}
In{\`e}s Ben El~Ouahma, Quentin~L. Meunier, Karine Heydemann, and Emmanuelle
  Encrenaz.
\newblock Side-channel robustness analysis of masked assembly codes using a
  symbolic approach.
\newblock {\em Journal of Cryptographic Engineering}, 9:231--242, 2019.

\bibitem[BKN22]{JCEng:BozKneNik22}
Dusan Bozilov, Miroslav Knezevic, and Ventzislav Nikov.
\newblock Optimized threshold implementations: securing cryptographic
  accelerators for low-energy and low-latency applications.
\newblock {\em Journal of Cryptographic Engineering}, 12(1):15--51, April 2022.

\bibitem[BWG{\etalchar{+}}22]{ProvableMaskingInRealWorld}
Arthur Beckers, Lennert Wouters, Benedikt Gierlichs, Bart Preneel, and Ingrid
  Verbauwhede.
\newblock Provable secure software masking in the real-world.
\newblock In Josep Balasch and Colin O'Flynn, editors, {\em Constructive
  Side-Channel Analysis and Secure Design}, pages 215--235, Cham, 2022.
  Springer International Publishing.

\bibitem[CAB25]{cassagne2025dalekunconventionalenergyawareheterogeneous}
Adrien Cassagne, Noé Amiot, and Manuel Bouyer.
\newblock Dalek: An unconventional and energy-aware heterogeneous cluster,
  2025.

\bibitem[CGD18]{COSADE:CorGroDin18}
Yann~Le Corre, Johann Gro{\ss}sch{\"a}dl, and Daniel Dinu.
\newblock Micro-architectural power simulator for leakage assessment of
  cryptographic software on {ARM} {Cortex}-{M3} processors.
\newblock In Junfeng Fan and Benedikt Gierlichs, editors, {\em COSADE 2018},
  volume 10815 of {\em {LNCS}}, pages 82--98. Springer, Cham, April 2018.

\bibitem[CJRR99]{C:CJRR99}
Suresh Chari, Charanjit~S. Jutla, Josyula~R. Rao, and Pankaj Rohatgi.
\newblock Towards sound approaches to counteract power-analysis attacks.
\newblock In Michael~J. Wiener, editor, {\em CRYPTO'99}, volume 1666 of {\em
  {LNCS}}, pages 398--412. Springer, Berlin, Heidelberg, August 1999.

\bibitem[CS20]{cassiers2020trivially}
Ga{\"e}tan Cassiers and Fran{\c{c}}ois-Xavier Standaert.
\newblock Trivially and efficiently composing masked gadgets with probe
  isolating non-interference.
\newblock {\em IEEE Transactions on Information Forensics and Security},
  15:2542--2555, 2020.

\bibitem[EGMP17]{AC:EGMP17}
Maik Ender, Samaneh Ghandali, Amir Moradi, and Christof Paar.
\newblock The first thorough side-channel hardware trojan.
\newblock In Tsuyoshi Takagi and Thomas Peyrin, editors, {\em ASIACRYPT~2017,
  Part~I}, volume 10624 of {\em {LNCS}}, pages 755--780. Springer, Cham,
  December 2017.

\bibitem[EWS14]{enumhassan}
Hassan Eldib, Chao Wang, and Patrick Schaumont.
\newblock Smt-based verification of software countermeasures against
  side-channel attacks.
\newblock In Erika {\'A}brah{\'a}m and Klaus Havelund, editors, {\em Tools and
  Algorithms for the Construction and Analysis of Systems}, pages 62--77,
  Berlin, Heidelberg, 2014. Springer Berlin Heidelberg.

\bibitem[FGM{\etalchar{+}}18]{TCHES:FGMPS18}
Sebastian Faust, Vincent Grosso, Santos {Merino Del Pozo}, Clara Paglialonga,
  and Fran{\c c}ois-Xavier Standaert.
\newblock Composable masking schemes in the presence of physical defaults {\&}
  the robust probing model.
\newblock {\em {IACR} {TCHES}}, 2018(3):89--120, 2018.

\bibitem[GHM22]{armistice}
Arnaud~de Grandmaison, Karine Heydemann, and Quentin~L. Meunier.
\newblock {ARMISTICE: Microarchitectural Leakage Modeling for Masked Software
  Formal Verification}.
\newblock {\em IEEE Transactions on Computer-Aided Design of Integrated
  Circuits and Systems}, 41(11):3733--3744, 2022.

\bibitem[GHP{\etalchar{+}}21]{USENIX:GHPMB21}
Barbara Gigerl, Vedad Hadzic, Robert Primas, Stefan Mangard, and Roderick
  Bloem.
\newblock Coco: Co-design and co-verification of masked software
  implementations on {CPUs}.
\newblock In Michael Bailey and Rachel Greenstadt, editors, {\em USENIX
  Security 2021}, pages 1469--1468. {USENIX} Association, August 2021.

\bibitem[GMK16]{grossDomainOrientedMaskingCompact2016}
Hannes Gross, Stefan Mangard, and Thomas Korak.
\newblock Domain-{{Oriented Masking}}: {{Compact Masked Hardware
  Implementations}} with {{Arbitrary Protection Order}}.
\newblock In {\em Proceedings of the 2016 {{ACM Workshop}} on {{Theory}} of
  {{Implementation Security}}}, pages 3--3. ACM, 2016.

\bibitem[GP99]{CHES:GouPat99}
Louis Goubin and Jacques Patarin.
\newblock {DES} and differential power analysis (the ``duplication'' method).
\newblock In {\c{C}etin Kaya}~Ko\c{c} and Christof Paar, editors, {\em
  CHES'99}, volume 1717 of {\em {LNCS}}, pages 158--172. Springer, Berlin,
  Heidelberg, August 1999.

\bibitem[GPM21]{AC:GigPriMan21}
Barbara Gigerl, Robert Primas, and Stefan Mangard.
\newblock Secure and efficient software masking on superscalar pipelined
  processors.
\newblock In Mehdi Tibouchi and Huaxiong Wang, editors, {\em ASIACRYPT~2021,
  Part~II}, volume 13091 of {\em {LNCS}}, pages 3--32. Springer, Cham, December
  2021.

\bibitem[GSM17]{keccakHO}
Hannes Gross, David Schaffenrath, and Stefan Mangard.
\newblock Higher-order side-channel protected implementations of keccak.
\newblock In {\em 2017 Euromicro Conference on Digital System Design (DSD)},
  pages 205--212, 2017.

\bibitem[GST{\etalchar{+}}17]{cv32e40p}
Michael Gautschi, Pasquale~Davide Schiavone, Andreas Traber, Igor Loi, Antonio
  Pullini, Davide Rossi, Eric Flamand, Frank~K. Gürkaynak, and Luca Benini.
\newblock Near-threshold risc-v core with dsp extensions for scalable iot
  endpoint devices.
\newblock {\em IEEE Transactions on Very Large Scale Integration (VLSI)
  Systems}, 25(10):2700--2713, 2017.

\bibitem[GXZ{\etalchar{+}}19]{qmverif}
Pengfei Gao, Hongyi Xie, Jun Zhang, Fu~Song, and Taolue Chen.
\newblock Quantitative verification of masked arithmetic programs against
  side-channel attacks.
\newblock In Tom{\'a}{\v{s}} Vojnar and Lijun Zhang, editors, {\em Tools and
  Algorithms for the Construction and Analysis of Systems}, pages 155--173,
  Cham, 2019. Springer International Publishing.

\bibitem[HB21]{CocoAlma}
Vedad Hadžić and Roderick Bloem.
\newblock {CocoAlma: A Versatile Masking Verifier}.
\newblock In {\em Proceedings of the 21st Conference on Formal Methods in
  Computer-Aided Design – FMCAD 2021}, volume~2 of {\em Conference Series:
  Formal Methods in Computer-Aided Design}, pages 14--23, Wien, 2021. TU Wien
  Academic Press.

\bibitem[HHB24]{closingthegap}
Johannes Haring, Vedad Hadži´c, and Roderick Bloem.
\newblock {Closing the Gap: Leakage Contracts for Processors with Transitions
  and Glitches}.
\newblock {\em IACR Transactions on Cryptographic Hardware and Embedded
  Systems}, 2024:110--132, 2024.

\bibitem[HOM06]{ACNS:HerOswMan06}
Christoph Herbst, Elisabeth Oswald, and Stefan Mangard.
\newblock An {AES} smart card implementation resistant to power analysis
  attacks.
\newblock In Jianying Zhou, Moti Yung, and Feng Bao, editors, {\em ACNS
  06International Conference on Applied Cryptography and Network Security},
  volume 3989 of {\em {LNCS}}, pages 239--252. Springer, Berlin, Heidelberg,
  June 2006.

\bibitem[ISW03]{C:IshSahWag03}
Yuval Ishai, Amit Sahai, and David Wagner.
\newblock Private circuits: Securing hardware against probing attacks.
\newblock In Dan Boneh, editor, {\em CRYPTO~2003}, volume 2729 of {\em {LNCS}},
  pages 463--481. Springer, Berlin, Heidelberg, August 2003.

\bibitem[KJJ99]{C:KocJafJun99}
Paul~C. Kocher, Joshua Jaffe, and Benjamin Jun.
\newblock Differential power analysis.
\newblock In Michael~J. Wiener, editor, {\em CRYPTO'99}, volume 1666 of {\em
  {LNCS}}, pages 388--397. Springer, Berlin, Heidelberg, August 1999.

\bibitem[Koc96]{C:Kocher96}
Paul~C. Kocher.
\newblock Timing attacks on implementations of {Diffie}-{Hellman}, {RSA},
  {DSS}, and other systems.
\newblock In Neal Koblitz, editor, {\em CRYPTO'96}, volume 1109 of {\em
  {LNCS}}, pages 104--113. Springer, Berlin, Heidelberg, August 1996.

\bibitem[KSM20]{AC:KniSasMor20}
David Knichel, Pascal Sasdrich, and Amir Moradi.
\newblock {SILVER} - statistical independence and leakage verification.
\newblock In Shiho Moriai and Huaxiong Wang, editors, {\em ASIACRYPT~2020,
  Part~I}, volume 12491 of {\em {LNCS}}, pages 787--816. Springer, Cham,
  December 2020.

\bibitem[MM22]{TCHES:MulMor22}
Nicolai M{\"u}ller and Amir Moradi.
\newblock {PROLEAD} {A} probing-based hardware leakage detection tool.
\newblock {\em {IACR} {TCHES}}, 2022(4):311--348, 2022.

\bibitem[MM24]{proleadv3}
Nicolai Müller and Amir Moradi.
\newblock {Robust but Relaxed Probing Model}.
\newblock {\em IACR Transactions on Cryptographic Hardware and Embedded
  Systems}, 2024:451--482, 2024.

\bibitem[MOP07]{Mangard2007PowerAA}
Stefan Mangard, Elisabeth Oswald, and Thomas Popp.
\newblock {\em {Power Analysis Attacks: Revealing the Secrets of Smart Cards}}.
\newblock Springer, 01 2007.

\bibitem[MPH23]{leakageverif}
Quentin~L. Meunier, Etienne Pons, and Karine Heydemann.
\newblock {LeakageVerif: Efficient and Scalable Formal Verification of Leakage
  in Symbolic Expressions}.
\newblock {\em IEEE Transactions on Software Engineering}, 49(6):3359--3375,
  2023.

\bibitem[MPW22]{TCHES:MarPagWeb22}
Ben Marshall, Dan Page, and James Webb.
\newblock {MIRACLE}: {MIcRo}-{ArChitectural} leakage evaluation {A} study of
  micro-architectural power leakage across many devices.
\newblock {\em {IACR} {TCHES}}, 2022(1):175--220, 2022.

\bibitem[MT23]{verifmsi}
Quentin Meunier and Abdul Taleb.
\newblock {VerifMSI: Practical Verification of Hardware and Software Masking
  Schemes Implementations}.
\newblock In {\em Proceedings of the 20th International Conference on Security
  and Cryptography - SECRYPT}, pages 520--527. INSTICC, SciTePress, 2023.

\bibitem[MWO16]{EPRINT:McCWhiOsw16}
David McCann, Carolyn Whitnall, and Elisabeth Oswald.
\newblock {ELMO}: Emulating leaks for the {ARM} {Cortex}-{M0} without access to
  a side channel lab.
\newblock Cryptology ePrint Archive, Report 2016/517, 2016.

\bibitem[NRR06]{ICICS:NikRecRij06}
Svetla Nikova, Christian Rechberger, and Vincent Rijmen.
\newblock Threshold implementations against side-channel attacks and glitches.
\newblock In Peng Ning, Sihan Qing, and Ninghui Li, editors, {\em ICICS 06},
  volume 4307 of {\em {LNCS}}, pages 529--545. Springer, Berlin, Heidelberg,
  December 2006.

\bibitem[OC14]{COSADE:OFlChe14}
Colin O'Flynn and Zhizhang~(David) Chen.
\newblock {ChipWhisperer}: An open-source platform for hardware embedded
  security research.
\newblock In Emmanuel Prouff, editor, {\em COSADE 2014}, volume 8622 of {\em
  {LNCS}}, pages 243--260. Springer, Cham, April 2014.

\bibitem[PMK{\etalchar{+}}11]{JC:PMKLWL11}
Axel Poschmann, Amir Moradi, Khoongming Khoo, Chu-Wee Lim, Huaxiong Wang, and
  San Ling.
\newblock Side-channel resistant crypto for less than 2,300 {GE}.
\newblock {\em Journal of Cryptology}, 24(2):322--345, April 2011.

\bibitem[RP10]{CHES:RivPro10}
Matthieu Rivain and Emmanuel Prouff.
\newblock Provably secure higher-order masking of {AES}.
\newblock In Stefan Mangard and Fran\c{c}ois-Xavier Standaert, editors, {\em
  CHES~2010}, volume 6225 of {\em {LNCS}}, pages 413--427. Springer, Berlin,
  Heidelberg, August 2010.

\bibitem[SCS{\etalchar{+}}21]{CCS:SCSWBY21}
Madura~A. Shelton, Lukasz Chmielewski, Niels Samwel, Markus Wagner, Lejla
  Batina, and Yuval Yarom.
\newblock Rosita++: Automatic higher-order leakage elimination from
  cryptographic code.
\newblock In Giovanni Vigna and Elaine Shi, editors, {\em ACM CCS 2021}, pages
  685--699. {ACM} Press, November 2021.

\bibitem[SM21]{TCHES:ShaMor21a}
Aein~Rezaei Shahmirzadi and Amir Moradi.
\newblock Second-order {SCA} security with almost no fresh randomness.
\newblock {\em {IACR} {TCHES}}, 2021(3):708--755, 2021.

\bibitem[SSB{\etalchar{+}}21]{NDSS:SSBRWY21}
Madura~A. Shelton, Niels Samwel, Lejla Batina, Francesco Regazzoni, Markus
  Wagner, and Yuval Yarom.
\newblock Rosita: Towards automatic elimination of power-analysis leakage in
  ciphers.
\newblock In {\em NDSS~2021}. The Internet Society, February 2021.

\bibitem[Wol13]{Yosys}
Claire Wolf.
\newblock Yosys open synthesis suite.
\newblock \url{https://yosyshq.net/yosys/}, 2013.

\bibitem[ZCF24]{prover}
Feng Zhou, Hua Chen, and Limin Fan.
\newblock {Prover - Toward More Efficient Formal Verification of Masking in
  Probing Model}.
\newblock {\em IACR Transactions on Cryptographic Hardware and Embedded
  Systems}, 2025(1):552–585, Dec. 2024.

\bibitem[ZMM23]{TCHES:ZeiMulMor23}
Jannik Zeitschner, Nicolai M{\"u}ller, and Amir Moradi.
\newblock {PROLEAD\_SW} probing-based software leakage detection for {ARM}
  binaries.
\newblock {\em {IACR} {TCHES}}, 2023(3):391--421, 2023.

\end{thebibliography}


\end{document}